\documentclass[nofootinbib, aps]{revtex4}

\usepackage{amstext,amsmath,amssymb}
\usepackage{amssymb,amsthm}
\usepackage{amsfonts}
\usepackage{graphicx}
\usepackage{latexsym}

\usepackage{psfrag}
\newcommand{\beq}{\begin{equation}}
\newcommand{\eeq}{\end{equation}}
\newcommand{\beqa}{\begin{eqnarray}}
\newcommand{\eeqa}{\end{eqnarray}}
\newcommand{\nn}{\nonumber}

\newcommand{\D}{\mathcal{D}}

\newcommand{\half}{\frac{1}{2}}

\newcommand{\SU}{\mathrm{SU}}

\newcommand{\tr}{\mathrm{tr}}

\begin{document}


\title{Wilson loops, geometric operators and fermions in 3d group field theory}
\author{{\bf R. Dowdall\footnote{richard.dowdall@maths.nottingham.ac.uk}}}
\vspace{0.5cm}
\affiliation{School of Mathematical Sciences,\\
University of Nottingham, \\ University Park,\\ Nottingham,\\ NG7 2RD, UK}

\begin{abstract}
Group field theories whose Feynman diagrams describe 3d gravity with a varying configuration of Wilson loop observables and 3d gravity with volume observables at each vertex are defined. The volume observables are created by the usual spin network grasping operators which require the introduction of vector fields on the group. We then use this to define group field theories that give a previously defined spin foam model for fermion fields coupled to gravity, and the simpler ``quenched'' approximation, by using tensor fields on the group. The group field theory naturally includes the sum over fermionic loops at each order of the perturbation theory.
\end{abstract}

\maketitle
\section{Introduction}
Group field theories (GFTs) provide a method for performing a perturbative sum over 2-complexes with weights that give a particular spin foam partition function evaluated on that 2-complex.  In this way, GFTs are a proposal for solving the problem of triangulation dependence in spin foam models and simultaneously provide a method for topology change \cite{rovelli,Oriti:2006se,Freidel:2005qe}.

An interesting problem in spin foam models for quantum gravity is the inclusion of matter.
Current proposals include point particles \cite{Freidel:2004vi} in three dimensions and ``strings'' \cite{Fairbairn:2007fb} in four dimensions.  These are motivated by the description of 3d particles as conical defects and result in a spin foam model with a deficit angle associated to each edge through which the particle propagates.  In an appropriate limit, the 3d amplitudes give a non-commutative field theory \cite{Freidel:2005bb}.
The formalism can be extended to include particles of arbitrary spin.
A group field theory for these particles was described in \cite{Oriti:2006jk} which reproduced Feynman diagrams for both gravity and matter.
The effective field theory can also be cast in the form of a GFT \cite{Fairbairn:2007sv} by perturbing the standard GFT for 3d gravity.

An alternative approach to adding matter in 3d spin foam models was given in \cite{Fairbairn:2006dn}.  The action for 3d gravity with a fermion field was discretised on a triangulation following usual spin foam techniques.  This resulted in modified vertex amplitudes for the spin foam model that contain an edge in the fundamental representation (representing the fermion) which then interacts with the geometry via a grasping operator. Unlike the particle model, the fermion model requires a sum over all the vacuum loops of the spin foam - as in lattice gauge theory.
In \cite{Fairbairn:2006dn}, it was observed that the fermion model was not topologically invariant since it now contains local degrees of freedom given by the fermions. A GFT would be one way to resolve this dependence on the triangulation.  In this paper we construct this GFT and we will see that it also naturally includes the sum over fermionic loops that appear in standard quantum field theory.

To arrive at the fermionic GFT, we first consider Wilson loops and volume operators since these are the building blocks of the model.
The Wilson loops require an extra argument in the fields to control the propagation of the loops and the volume operators will require the use of higher spin fields due to the nature of the operators involved.
For illustrative purposes, we also include the ``quenched'' fermion model, which neglects half of the loop configurations, as this contains all of the main features of the theory without having to include more complicated configurations of loops.  We then describe the full GFT for the fermion model and briefly comment on its similarities and differences to the GFT for particles coupled to the Ponzano-Regge model.

\section{Wilson loops}
\label{wilson section}
We start by defining the GFT for 3d Euclidean gravity with Wilson loops.  The Feynman diagrams of this field theory will give Ponzano-Regge spin foam models with Wilson loop observables \cite{Freidel:2004nb}.  In fact, the GFT will give all possible Wilson loops in the spin foam so in itself is unlikely to be physically interesting.  However, it will be useful for defining the matter coupling.
For the Ponzano-Regge model defined on a triangulation $\Delta$, the Wilson loop observable $\mathcal{O}_{\mathcal{W}}^J$ is defined as the trace in the representation $J$ of the holonomy around a closed loop $\mathcal{W}$ in the dual triangulation $\Delta*$
\beq
\mathcal{O}_{\mathcal{W}}^J = \tr^J( \prod^{\rightarrow}_{e^* \in \mathcal{W}} g_{e^*} ).
\eeq
The expectation value of $\mathcal{O}_{\mathcal{W}}^J$ is then given by
\beq
\langle \mathcal{O}_{\mathcal{W}}^J  \rangle_{\mbox{\tiny PR}}(\Delta) = \int_{SU(2)} \prod_{e^*} dg_{e^*}  \
\mathcal{O}_{\mathcal{W}}^J \
\prod_{f^*}   \delta \left( G_{f^*}     \right).
\eeq
The group variables $g_{e^*}$ represent $\SU(2)$ holonomies along dual edges $e^*$ of $\Delta$, $G_{f*} = \prod^{\rightarrow}_{e* \in f*} g_{e*}$ is the ordered product of holonomies around a dual face $f^*$ of $\Delta$. Instead of the usual 6j symbol, the spin network vertex amplitudes for the model will contain an extra edge in the spin $J$ representation at any vertex that is traversed by the Wilson loop.
\begin{center}
\psfrag{j1}{$j_1$}
\psfrag{j2}{$j_2$}
\psfrag{j3}{$j_3$}
\psfrag{j4}{$j_4$}
\psfrag{j5}{$j_5$}
\psfrag{j6}{$j_6$}
\psfrag{J}{$J$}
\includegraphics[scale=0.45]{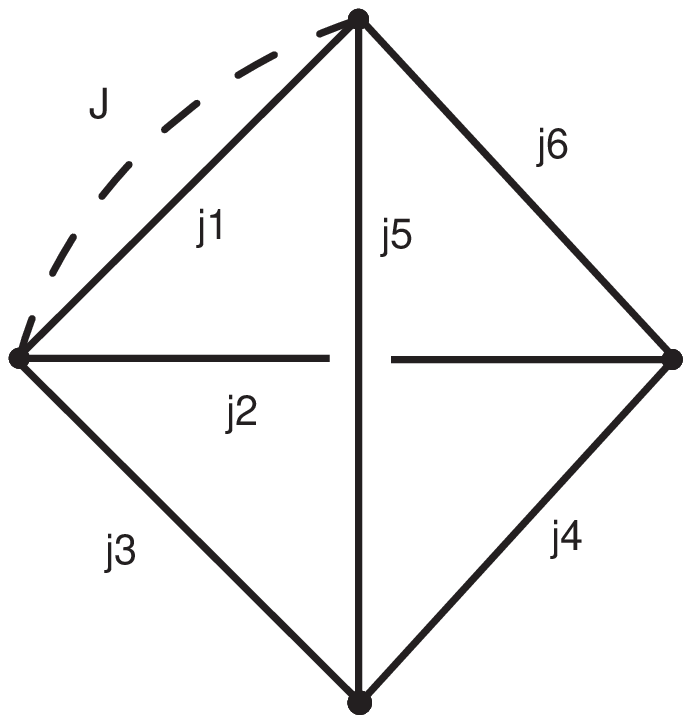}
\end{center}
We can now define the GFT for a Wilson loop observable.  Since the GFT sums over 2-complexes, we will in fact create all possible Wilson loops at each value of the GFT expansion parameter.  This will include disjoint loops so from now on we use $ \tilde{ \mathcal{W} }$ to denote a configuration of any number of Wilson loops.  Unless the loop configuration $\tilde{ \mathcal{W}}$ completely saturates the spin foam, we will also require vertex amplitudes for those vertices that are not part of the loop. For this, we recall the pure gravity field \cite{Boulatov:1992vp} defined as a map from three copies of SU(2) to the complex numbers
\beq
\label{GR field 1}
\phi(g_1,g_2,g_3) : SU(2) \times SU(2) \times SU(2) \rightarrow \mathbb{C}
\eeq
As usual, we require that this is invariant under permutations $\sigma$ of the group variables
\beq
\phi(g_1,g_2,g_3) = \phi(g_{\sigma(1)},g_{\sigma(2)},g_{\sigma(3)})
\eeq
and we make the field $SU(2)$ invariant
\beq
\phi(g g_1,g g_2,g g_3) = \phi(g_1,g_2,g_3) \ \ ; \ g \in \SU(2)
\eeq
with the projector $P_\alpha$
\beq
P_\alpha \phi(g_1,g_2,g_3) = \int_{SU(2)}  d\alpha \ \phi(\alpha g_1,\alpha g_2,\alpha g_3)
\eeq
With the field defined as above, we can write the action $S_{\mbox{\tiny GR}}$ for the GFT whose Feynman diagrams produce Ponzano-Regge spin foams.
\beqa
S_{\mbox{\tiny GR}}[\phi , \lambda_{\mbox{\tiny GR}} ]
&=&
\frac{1}{2} \int \prod^3_{i=1} dg_i
P_{\alpha_1}  \phi(g_1,g_2,g_3)
P_{\alpha_2} \phi(g_1,g_2,g_3) \nn \\
&+&
\frac{\lambda_{\mbox{\tiny GR}}}{4} \int \prod^6_{i=1} dg_i
P_{\alpha_1} \phi(g_1,g_2,g_3)
P_{\alpha_2} \phi(g_1,g_5,g_6)
P_{\alpha_3} \phi(g_2,g_4,g_6)
P_{\alpha_4} \phi(g_3,g_4,g_5)
\eeqa
To include the Wilson loops, we define an additional field $\psi$ with an extra argument
\beq
\psi(g_1,g_2,g_3 ; g) : SU(2) \times SU(2) \times SU(2) \times SU(2) \rightarrow \mathbb{C}
\eeq
This extra variable $g$ will control the way the Wilson loop propagates throughout the 2-complex.
We project the additional argument of the field onto the representation $J$ specified by the Wilson loop with the following projector
\beq
\psi^J(g_1,g_2,g_3 ; g) = (P^J \psi)(g_1,g_2,g_3 ; g) = \int_{SU(2)} dh \ d_J \chi^J(g h^{-1}) \psi(g_1,g_2,g_3 ; h)
\eeq
where $d_J=2J+1$ is the dimension of the irreducible representation $J$ and $\chi^J$ is the character.
We also demand that the field is $SU(2)$ invariant by projecting
\beq
P_\alpha \psi(g_1,g_2,g_3;g) = \int_{SU(2)}  d\alpha \ \psi( g_1 \alpha, g_2 \alpha, g_3 \alpha ;  g \alpha),
\eeq
this will create the four-valent intertwiners in the Wilson loop vertex amplitudes. Finally, we demand permutation symmetry on the first three arguments.
\beq
 \psi(g_1,g_2,g_3;g) = \psi(g_{\sigma(1)},g_{\sigma(2)},g_{\sigma(3)} ; g )
\eeq
We can use the Peter-Weyl theorem to perform a Fourier decomposition of the fields
\beqa
P_\alpha \psi^J(g_1,g_2,g_3;g) &=&
\sum_{     \substack{  j_i , m_i, n_i, k_i, a ,b      \\    1 \leq i \leq 4       } }
\int_{SU(2)} d\alpha \int_{SU(2)} dh \
\psi^{j_1 j_2 j_3 j_4}_{m_1 k_1 m_2 k_2 m_3 k_3 m_4 k_4}
\sqrt{d_{j_1} d_{j_2}  d_{j_3} d_{j_4}}
\nn \\ && \times
D^{j_1}_{m_1 n_1}(g_1) D^{j_2}_{m_2 n_2}(g_2) D^{j_3}_{m_3 n_3}(g_3) D^{j_4}_{m_4 n_4}(h)
\nn \\ && \times
\ D^{j_1}_{n_1 k_1}(\alpha) D^{j_2}_{n_2 k_2}(\alpha) D^{j_3}_{n_3 k_3}(\alpha) D^{j_4}_{n_4 k_4}(\alpha)
\nn \\ && \times
 \ D^{J}_{a b}(g) D^{J}_{ba}(h^{-1})
\eeqa
Performing the group integrals reduces this to
\beqa
P_\alpha \psi^J(g_1,g_2,g_3;g) &=&
\sum_{\substack{j_i , m_i, n_i, k_i , s \\ 1 \leq i \leq 4}}
\psi^{j_1 j_2 j_3 J s}_{m_1  m_2  m_3 m_4 } \sqrt{d_{j_1} d_{j_2}  d_{j_3} d_{j_4}}
\nn \\ && \times
D^{j_1}_{m_1 n_1}(g_1) D^{j_2}_{m_2 n_2}(g_2) D^{j_3}_{m_3 n_3}(g_3) D^{J}_{m_4 n_4}(g)
C^{j_1 j_2 j_3 J s }_{n_1 n_2 n_3 n_4}
\eeqa
Where $C^{j_1 j_2 j_3 J \ s }_{k_1 k_2 k_3 J}$ is a four valent $SU(2)$ intertwiner labelled by $s$ and we have redefined the coefficients as
\beq
\psi^{j_1 j_2 j_3 J s}_{m_1  m_2  m_3 m_4 } =
\psi^{j_1 j_2 j_3 J}_{m_1 k_1 m_2 k_2 m_3 k_3 m_4 k_4}
C^{j_1 j_2 j_3 J s}_{k_1 k_2 k_3 k_4}
\eeq
We define the action as a functional of the two fields to be
\beqa
S_{\mbox{\tiny W}}[\phi , \psi^J ,\lambda_{\mbox{\tiny GR}}, \lambda_{\mbox{\tiny W}} ]
&=& S_{\mbox{{\tiny GR}}}[\phi , \lambda_{\mbox{\tiny GR}} ]
+ \frac{1}{2} \int \prod^3_{i=1} dg_i P_\alpha \psi^J(g_1,g_2,g_3 ; g) P_\alpha \psi^J(g_1,g_2,g_3 ; g) \nn \\
&+&  \frac{\lambda_{\mbox{\tiny W}}}{4} \int \prod^6_{i=1} dg_i \ dg P_{\alpha_1} \psi^J(g_1,g_2,g_3 ; g) P_{\alpha_2} \psi^J(g_1,g_5,g_6 ; g) P_{\alpha_3} \phi(g_2,g_4,g_6) P_{\alpha_4} \phi(g_3,g_4,g_5)          \nn \\
\eeqa
With this action  we can now compute the propagators and vertex amplitudes of the theory.

\subsection{Feynman amplitudes}

We can now consider the partition function for the theory and express it in terms of the Feynman diagram expansion.  The expansion will implement a sum over 2-complexes and Wilson loop configurations on each 2-complex.
\beq
Z = \int  \mathcal{D} \psi^J   \D \phi e^{- S_{\mbox{\tiny W}}[\phi, \psi^J, \lambda_{\mbox{\tiny GR}}, \lambda_{\mbox{\tiny W}}]}
=\sum_{\Gamma}  \frac{ \lambda_{\mbox{\tiny GR}}^{v_{\mbox{\tiny GR}}[\Gamma ]}   \lambda_{\mbox{\tiny W}}^{v_{\mbox{\tiny W}}[\Gamma ]}  }{\mathrm{sym}(\Gamma )} Z[\Gamma].
\eeq
 Here $\Gamma$ denotes a particular Feynman diagram in this sum.  The number of pure gravity vertices in $\Gamma$ is denoted $v_{\mbox{\tiny GR}}[\Gamma ]$, the number of Wilson loop vertices is $v_{\mbox{\tiny W}}[\Gamma ]$ and the symmetry factor of the diagram is written as $\mathrm{sym}(\Gamma )$.
Each term $Z[\Gamma ]$ in the sum gives the spin foam partition function evaluated on the 2-complex defined by $\Gamma$.

The Feynman rules for the GFT are as follows.  There are two propagators, the pure gravity propagator and one with an extra delta function that determines the path of the Wilson loop.
\begin{center}
\begin{tabular}{c}
\psfrag{g1}{$g_1$}
\psfrag{g2}{$g_2$}
\psfrag{g3}{$g_3$}
\includegraphics[scale=0.35]{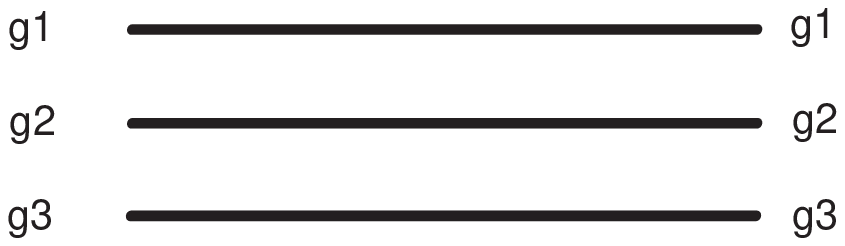}
\end{tabular}
, \ \ \ \ \ \ \ \ \ \ \ \ \ \
\begin{tabular}{c}
\psfrag{g1}{$g_1$}
\psfrag{g2}{$g_2$}
\psfrag{g3}{$g_3$}
\psfrag{g}{$g$}
\includegraphics[scale=0.35]{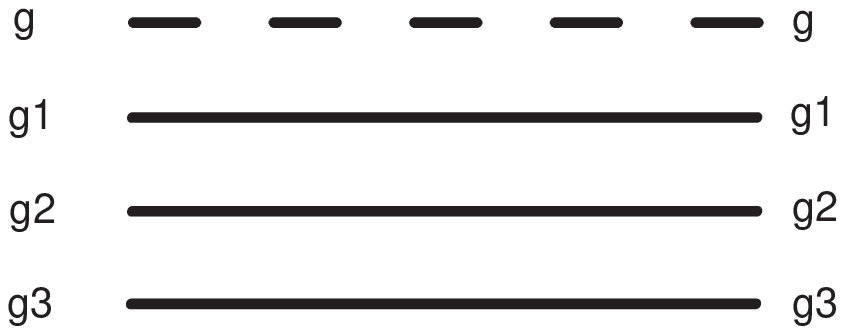}
\end{tabular}
\end{center}
Similarly, there are two vertex amplitudes corresponding to a vertex that is not intersected by a loop and one that is.
\begin{center}
\psfrag{g1}{$g_1$}
\psfrag{g2}{$g_2$}
\psfrag{g3}{$g_3$}
\psfrag{g4}{$g_4$}
\psfrag{g5}{$g_5$}
\psfrag{g6}{$g_6$}
\psfrag{g}{$g$}
 \begin{tabular}{c}
 \includegraphics[scale=0.45]{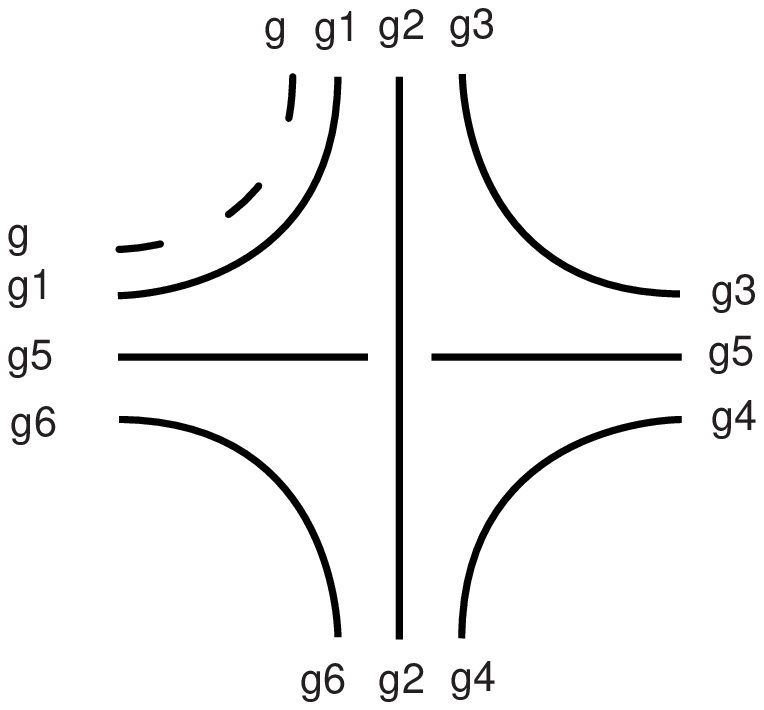}
 \end{tabular}
 { \Huge $\Rightarrow$}
 \begin{tabular}{c}
 \psfrag{j1}{$j_1$}
\psfrag{j2}{$j_2$}
\psfrag{j3}{$j_3$}
\psfrag{j4}{$j_4$}
\psfrag{j5}{$j_5$}
\psfrag{j6}{$j_6$}
\psfrag{J}{$J$}
\includegraphics[scale=0.45]{wilsonamp.eps}
 \end{tabular}\\
 \begin{tabular}{c}
 \psfrag{g1}{$g_1$}
\psfrag{g2}{$g_2$}
\psfrag{g3}{$g_3$}
\psfrag{g4}{$g_4$}
\psfrag{g5}{$g_5$}
\psfrag{g6}{$g_6$}
  \includegraphics[scale=0.45]{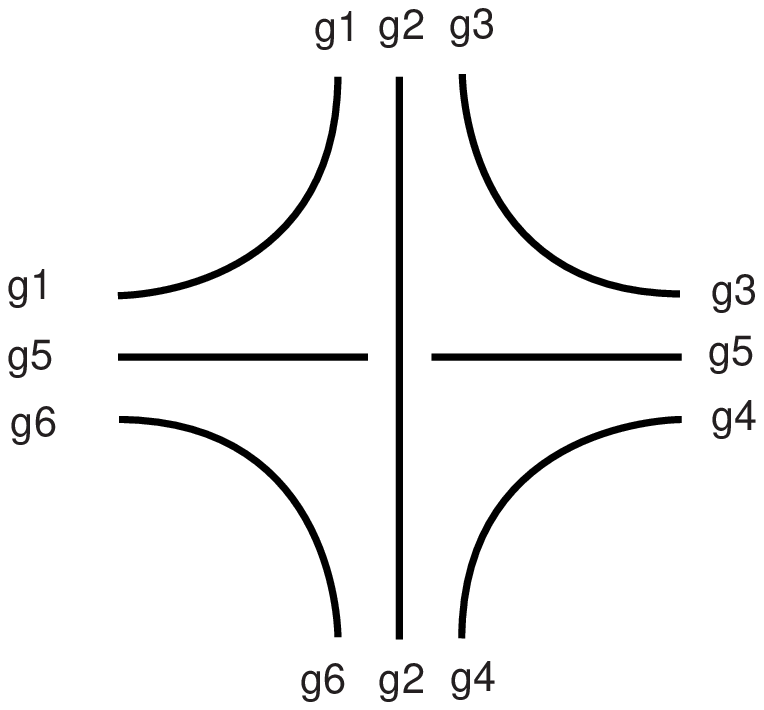}
 \end{tabular} { \Huge $\Rightarrow$}
 \begin{tabular}{c}
  \psfrag{j1}{$j_1$}
\psfrag{j2}{$j_2$}
\psfrag{j3}{$j_3$}
\psfrag{j4}{$j_4$}
\psfrag{j5}{$j_5$}
\psfrag{j6}{$j_6$}
\psfrag{J}{$J$}
 \includegraphics[scale=0.45]{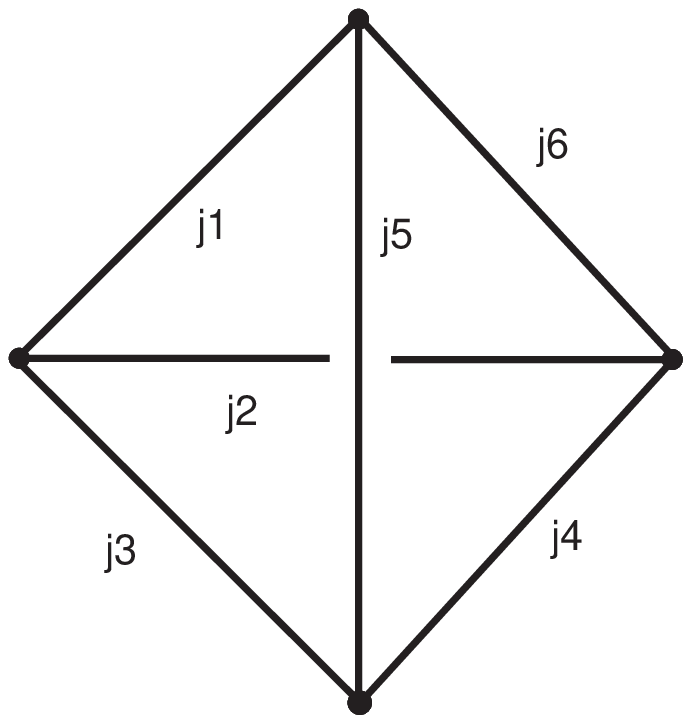}
 \end{tabular}
\end{center}
Note that for a fixed $v_{\mbox{\tiny GR}}$, $v_{\mbox{\tiny W}}$ is the number of dual vertices traversed by the Wilson loops.  Special cases include $v_{\mbox{\tiny W}} = 0$ which corresponds to a pure partition function without observables and $v_{\mbox{\tiny GR}}=0$ which contains all possible Wilson loops that completely saturate the spin foam.

\section{Geometric operators}
\label{volume section}
The geometric operators we consider here are constructed from polynomials of the discrete frame field used in the construction of 3d spin foams.  Expectation values of these operators can be computed using the technique of generating functionals \cite{Freidel:1998pt} and result in so called grasping operators acting on the spin network vertex amplitudes.  In particular, we will consider the volume operator although others can be constructed in a similar fashion.  There are different ways in which to construct a quantum mechanical volume operator in terms of the frame fields which are equivalent classically.  These have been studied in detail in \cite{Hackett:2006gp}.  For a tetrahedron, $t$, we define the volume operator ${\mathcal V_t}$ to be
\beq
\label{volume1}
{\mathcal V_t} = \sum_{e_1,e_2,e_3} \epsilon_{IJK} \, X_{e_1}^I X_{e_2}^J X_{e_3}^K.
\eeq
The $X_e^I$ are the discrete frame fields associated to each edge $e$ of $\Delta$ and carry an $\SU (2)$ Lie algebra index.  The summation is over all triples of edges $e_1,e_2,e_3$ in $t$ that give a non-zero volume.
To compute expectation values of this type of observable, one introduces ${\mathfrak su}(2)$ valued sources  $J_{e}$ on each edge $e$ and defines the generating functional $Z[J_{e}]$ to be
\beq
\label{genBF}
Z[J_{e}] = \prod_{e^*} \int_{\SU(2)} dg_{e^*} \; \prod_{e'} \int_{{\mathfrak su}(2)}dX_{e'}
\; \exp \left( {{i} \sum_e {\rm Tr} \, X_e \left(G_e + J_e \right)} \right).
\eeq
The (unnormalised) expectation value of the volume operator can then be computed using
\beqa
\label{exV}
\langle {\cal V}_t \rangle_{\mbox{\tiny GR}}(\Delta) &=&
\prod_{e^*} \int_{\SU(2)} dg_{e^*} \; \prod_{e'} \int_{{\mathfrak su}(2)}dX_{e'}
\; \sum_{e_1,e_2,e_3} \epsilon_{IJK}X^I_{e_1}X^J_{e_2}X^K_{e_3} \;
\exp \left( {{i} \sum_e {\rm Tr} \, X_e G_e } \right)
\nn \\
&=& i
 \left(\sum_{e_1,e_2,e_3} \epsilon_{IJK}\frac{\delta}{\delta J_{e_1}^I}
\frac{\delta}{\delta J_{e_2}^J} \frac{\delta}{\delta J_{e_3}^K}\right)\;Z[J_{e}]\;\Big|_{J_{e}=0}.
\eeqa
The source derivations insert Lie algebra generators into the appropriate edges which are then contracted with $\epsilon_{IJK}$. Remembering that $\epsilon_{IJK}$ is an intertwiner between three vector representations, then, up to some normalisation factors, the resulting spin network vertex amplitudes will be of the form
\begin{center}
\begin{tabular}{c}
  \psfrag{j1}{$j_1$}
\psfrag{j2}{$j_2$}
\psfrag{j3}{$j_3$}
\psfrag{j4}{$j_4$}
\psfrag{j5}{$j_5$}
\psfrag{j6}{$j_6$}
\psfrag{1}{$1$}
 \includegraphics[scale=0.45]{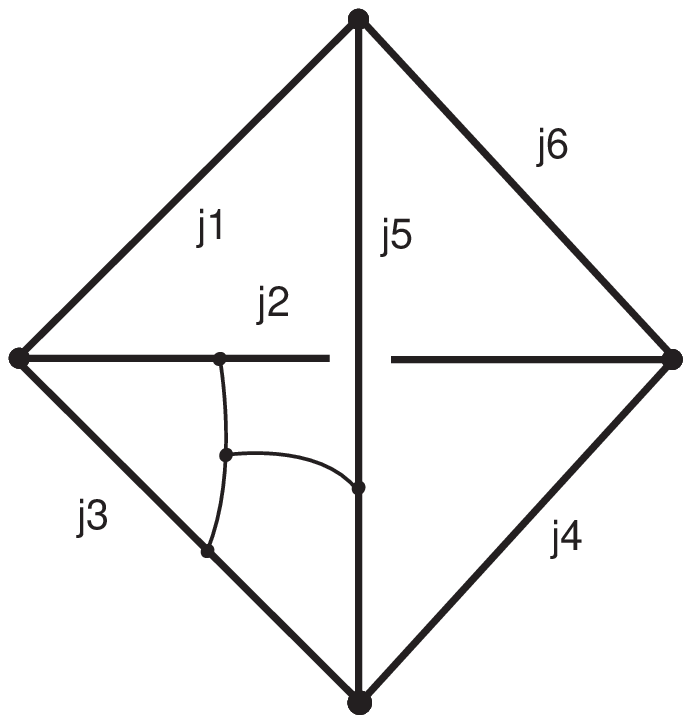}
 \end{tabular}
\end{center}

\paragraph*{\bf The GFT: }
We begin by observing that the grasping operation cannot be implemented simply as a projector on the standard Boulatov GFT as the grasping would be non-local in the fields. We resolve this problem by considering vector fields on $\SU(2)$.  With an appropriate invariance property, this will allow us to construct the graspings by contracting with the index of the field.
We begin by defining a vector field on three copies of $\SU(2)$
\beq
\label{grasping field}
\phi_a(g_1,g_2,g_3) : \SU(2) \times \SU(2) \times \SU(2) \rightarrow {\mathbb{V}}_1
\eeq
Where $\mathbb{V}_1$ is the $j=1$ $\SU(2)$ representation space and $a$ is an index with values $1,2,3$.
In order to create the necessary intertwiners, we use the following projector on the field
\beq
\tilde{P}_{\alpha ,\beta} \phi_a(g_1,g_2,  g_3) = \int_{SU(2)}d \alpha \  d\beta \ D^1_{ab}(\beta) \phi_b( g_1 \alpha ,g_2 \alpha , g_3 \beta \alpha^{-1} \beta).
\eeq
Note that with this projection, the field satisfies the following $SU(2)$ invariance property
\beq
D^1_{ab}(h) \tilde{P}_{\alpha ,\beta} \phi_b(g_1 h ,g_2 h ,g_3 h) =
 \tilde{P}_{\alpha ,\beta} \phi_a( g_1  ,g_2  , g_3 ).
\eeq
We also require the invariance under permutations $\sigma$ of the group variables
by summing over permutations of the field arguments
\beq
\phi_a(g_1,g_2,g_3) = \sum_{\sigma} \phi_a(g_{\sigma(1)},g_{\sigma(2)},g_{\sigma(3)}).
\eeq
The mode expansion of this field is then
\beqa
\tilde{P}_{\alpha ,\beta} \phi_a(g_1,g_2,  g_3) &=&
\sum_{\substack{j_i , m_i, n_i, k_i, p_3, q_3 a ,b \\ 1 \leq i \leq 3}}
\phi^{j_1 j_2 j_3 1}_{m_1 k_1 m_2 k_2 m_3 k_3  b}
\sqrt{d_{j_1} d_{j_2}  d_{j_3} d_{j_4}}
D^{j_1}_{m_1 n_1}(g_1) D^{j_2}_{m_2 n_2}(g_2) D^{j_3}_{m_3 n_3}(g_3)
\nn \\ && \times
\int_{SU(2)} d\alpha \ D^{j_1}_{n_1 k_1}(\alpha) D^{j_2}_{n_2 k_2}(\alpha) D^{j_3}_{p_3 q_3}(\alpha^{-1} )
\int_{SU(2)} d\beta \ D^{j_3}_{n_3p_3}(\beta) D^{j_3}_{q_3 k_3}(\beta) D^{j_3}_{ab}(\beta).
\eeqa
Which reduces to
\beqa
\tilde{P}_{\alpha ,\beta} \phi_a(g_1,g_2,  g_3) &=&
\sum_{\substack{j_i , m_i, n_i, k_i,  q_3 a ,b \\ 1 \leq i \leq 3}}
\phi^{j_1 j_2 j_3  }_{m_1  m_2  m_3  } \sqrt{d_{j_1} d_{j_2}  d_{j_3} d_{j_4}}
C^{j_1 j_2 j_3}_{n_1 n_2  q_3}  C^{j_3 j_3 1}_{n_3 q_3 a}
\nn \\ && \times
D^{j_1}_{m_1 n_1}(g_1) D^{j_2}_{m_2 n_2}(g_2) D^{j_3}_{m_3 n_3}(g_3),
\eeqa
where the modes have been redefined as
\beqa
\phi^{j_1 j_2 j_3  }_{m_1  m_2  m_3  } =
\sum_{\substack{ n_i, k_i, p_3 ,b \\ 1 \leq i \leq 3}}
\phi^{j_1 j_2 j_3 1}_{m_1 k_1 m_2 k_2 m_3 k_3  b}
C^{j_1 j_2 j_3}_{k_1 k_2 p_3} C^{j_3 j_3 1}_{p_3 k_3 b}
\eeqa
With this new field, we define the action for the GFT that has a volume grasping on every vertex amplitude.
\beqa
S_{\mbox{\tiny VOL}}[\phi , \lambda_{\mbox{\tiny VOL}} ]
&=&
\frac{1}{2} \int \prod^3_{i=1} dg_i
P_{\alpha_1}  \phi(g_1,g_2,g_3)
P_{\alpha_2} \phi(g_1,g_2,g_3)
\nn \\ & & +
\frac{1}{2} \int \prod^3_{i=1} dg_i   \tilde{P}_{\alpha_1 ,\beta_1} \phi_a(g_1,g_2,g_3)\tilde{P}_{\alpha_2 ,\beta_2} \phi_b(g_1,g_2,g_3) \delta_{ab}
\nn \\
&& + \frac{\lambda_{\mbox{\tiny VOL}}}{4} \int \prod^6_{i=1} dg_i \tilde{P}_{\alpha_1 ,\beta_1} \phi_a(g_1,g_2,g_3)  \tilde{P}_{\alpha_2 ,\beta_2} \phi_b(g_1,g_5,g_6)
\tilde{P}_{\alpha_3 ,\beta_3} \phi_c(g_2,g_4,g_6) P_{\alpha_4} \phi(g_3,g_4,g_5)  \ \epsilon_{abc}
\eeqa

\paragraph*{\bf Feynman rules:}

The partition function and its Feynman diagram expansion is
\beq
Z = \int  \mathcal{D} \phi_a   \D \phi e^{- S_{\mbox{\tiny VOL}}[\phi, \phi_a, \lambda_{\mbox{\tiny VOL}}]}
=\sum_{\Gamma}  \frac{ \lambda_{\mbox{\tiny VOL}}^{v_{\mbox{\tiny VOL}}[\Gamma ]}   }{\mathrm{sym}(\Gamma )} Z[\Gamma].
\eeq
With $v_{\mbox{\tiny VOL}}[\Gamma ]$ giving the number of vertices of $\Gamma$.
There are two propagators for the theory, one for the pure gravity field and one between the grasped fields.
\begin{center}
\begin{tabular}{c}
\psfrag{g1}{$g_1$}
\psfrag{g2}{$g_2$}
\psfrag{g3}{$g_3$}
\includegraphics[scale=0.35]{grpropagator}
\end{tabular}
, \ \ \ \ \ \ \ \ \
\begin{tabular}{c}
\psfrag{g1}{$g_1$}
\psfrag{g2}{$g_2$}
\psfrag{g3}{$g_3$}
\includegraphics[scale=0.35]{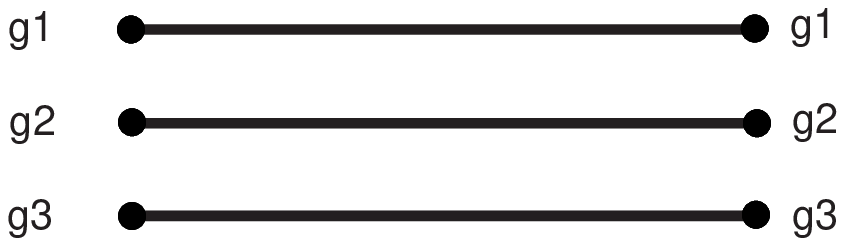}
\end{tabular}
\end{center}
There is a single vertex that is formed by the interaction of three grasping fields with one pure gravity.  The dotted lines connect with the vector field propagator while the gravity fields use the usual propagator, note that this means the vector field cannot propagate to the gravity field.
\begin{center}
\begin{tabular}{c}
\psfrag{g1}{$g_1$}
\psfrag{g2}{$g_2$}
\psfrag{g3}{$g_3$}
\psfrag{g4}{$g_4$}
\psfrag{g5}{$g_5$}
\psfrag{g6}{$g_6$}
 \includegraphics[scale=0.40]{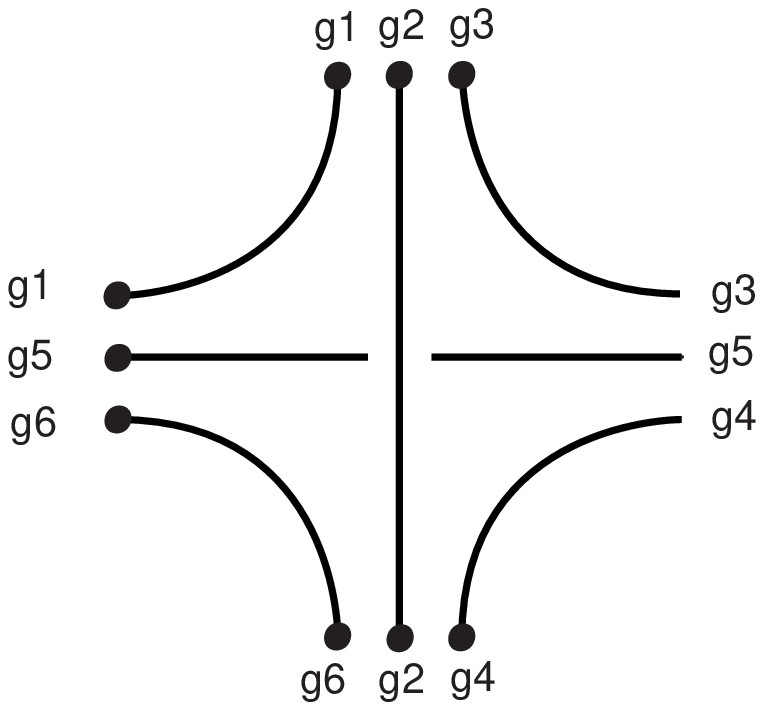}
 \end{tabular}
 { \Huge $\Rightarrow$}
 \begin{tabular}{c}
  \psfrag{j1}{$j_1$}
\psfrag{j2}{$j_2$}
\psfrag{j3}{$j_3$}
\psfrag{j4}{$j_4$}
\psfrag{j5}{$j_5$}
\psfrag{j6}{$j_6$}
\psfrag{1}{$1$}
 \includegraphics[scale=0.45]{graspamp}
 \end{tabular}
\end{center}

Note that due to the form of the vertex, the Feynman diagram expansion will not produce all 2-complexes but a subset of them. We ignore this issue as we are only considering $S_{\mbox{\tiny VOL}}$ as a stepping stone to construction of the fermionic model.  Since we have imposed the permutation symmetry on the field, we will obtain all possible graspings on the tetrahedral spin network.  These different graspings are described in \cite{Hackett:2006gp}.

\section{Fermions}

Using the techniques above, it will now be possible to write the GFT whose Feynman diagrams give fermion fields coupled to 3d spin foam gravity.  First, we briefly recall the fermion model and describe how to obtain the massless case, for more details refer to \cite{Fairbairn:2006dn}.

\subsection{Massless Fermions in 3d gravity}

The fermion model is obtained by discretising the gravity sector as usual and then placing the fermion fields at the dual vertices $v^*$ of $\Delta$ in the same way as conventional lattice gauge theory.  Once the fermionic integration has been performed, one is left with a sum over two different sets of loop $\ell_1 , \ell_2$ that saturate all the dual vertices of $\Delta$.  As for the Wilson loops, these can consist of more than one loop so we refer to $\mathcal{L}^{n_1}_1 , \mathcal{L}^{n_2}_2$ as loop configurations containing $n_i$ loops as $\mathcal{L}_i^{n_i} = \{\ell_1 , ... \ell_{n_i}  \} , i=1,2$.  For each loop $l$, we insert the following into the pure gravity partition function
\beq
\label{F loop}
F_{\ell}=\tr  \left(  \prod_{ e^* \in \ell  } \left( A_{e^*}  D^{\frac{1}{2}}(g_{e^*})  -   D^{\frac{1}{2}}(g_{e^*}) A_{{e^*}^{-1}}   \right)             \right)
\eeq
Where $D^{\frac{1}{2}}(g)$ is the representation matrix of $g$ in the fundamental representation and the spinor indices have been suppressed.  The $A_{e*}$ term describes the area of the triangle $f$ dual to $e^*$ through which the fermionic loop $l$ propagates and is given by
\beq
A_{e^*} = \sum_{e_1 , e_2 \in f}  \epsilon_{IJK} \sigma^I X_{e_1}^J  X_{e_2}^K.
\eeq
Here, the summation is over edges $e$ in the triangular face $f$ dual to $e^*$ and $\sigma^I$ denotes the Pauli matrices considered as an intertwiner between two spinor and one vector representation space.

In fact, the spin foam model in \cite{Fairbairn:2006dn} considered massive fermions and the integration over the Grassmann valued fermion fields was performed by using a hopping parameter expansion in the inverse fermion mass  to calculate the Pfaffian of the Dirac matrix defining the action (equation (43) of \cite{Fairbairn:2006dn}). The massless case can be obtained directly from the Grassmann integration, i.e. by repeated use of equation (46) of \cite{Fairbairn:2006dn} without the volume term, and one obtains the sum over loop configurations given above. Alternatively, using the hopping parameter expansion, if one considers the limit in which the mass parameter is small then the highest order term in the expansion dominates. This is the term $\Gamma_n$ in equation (50) of \cite{Fairbairn:2006dn} which again gives the sum over loop configurations described above. There are two different configurations of loops in the spin foam since it is actually the Pfaffian squared which appears in the path integral.

\begin{figure}
\begin{tabular}{ccc}
  \psfrag{j1}{$j_1$}
\psfrag{j2}{$j_2$}
\psfrag{j3}{$j_3$}
\psfrag{j4}{$j_4$}
\psfrag{j5}{$j_5$}
\psfrag{j6}{$j_6$}
  \includegraphics[scale=0.35]{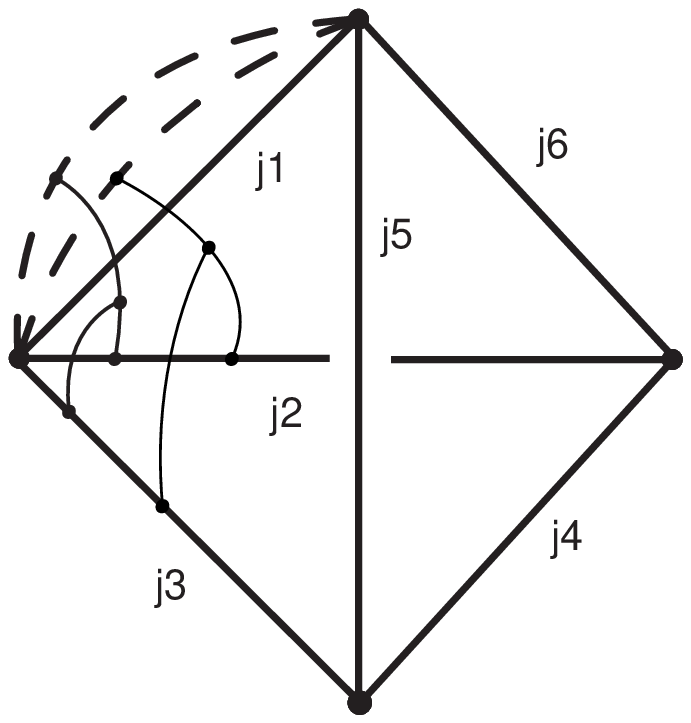} &
  \psfrag{j1}{$j_1$}
\psfrag{j2}{$j_2$}
\psfrag{j3}{$j_3$}
\psfrag{j4}{$j_4$}
\psfrag{j5}{$j_5$}
\psfrag{j6}{$j_6$}
\includegraphics[scale=0.35]{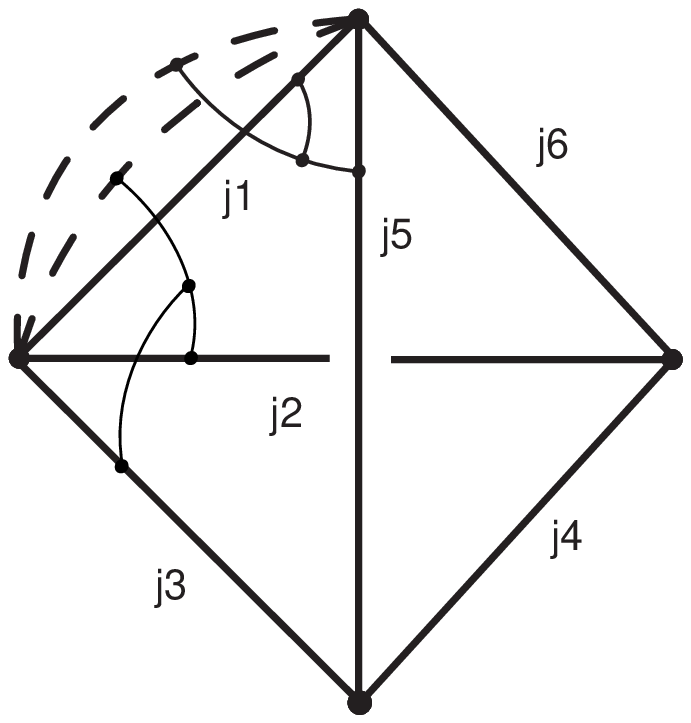} &
\psfrag{j1}{$j_1$}
\psfrag{j2}{$j_2$}
\psfrag{j3}{$j_3$}
\psfrag{j4}{$j_4$}
\psfrag{j5}{$j_5$}
\psfrag{j6}{$j_6$}
\includegraphics[scale=0.35]{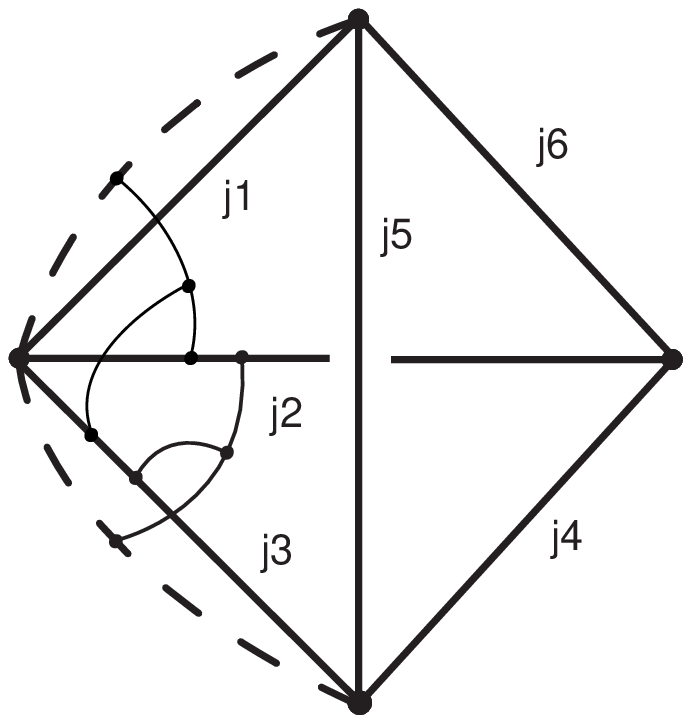} \\
\psfrag{j1}{$j_1$}
\psfrag{j2}{$j_2$}
\psfrag{j3}{$j_3$}
\psfrag{j4}{$j_4$}
\psfrag{j5}{$j_5$}
\psfrag{j6}{$j_6$}
\includegraphics[scale=0.35]{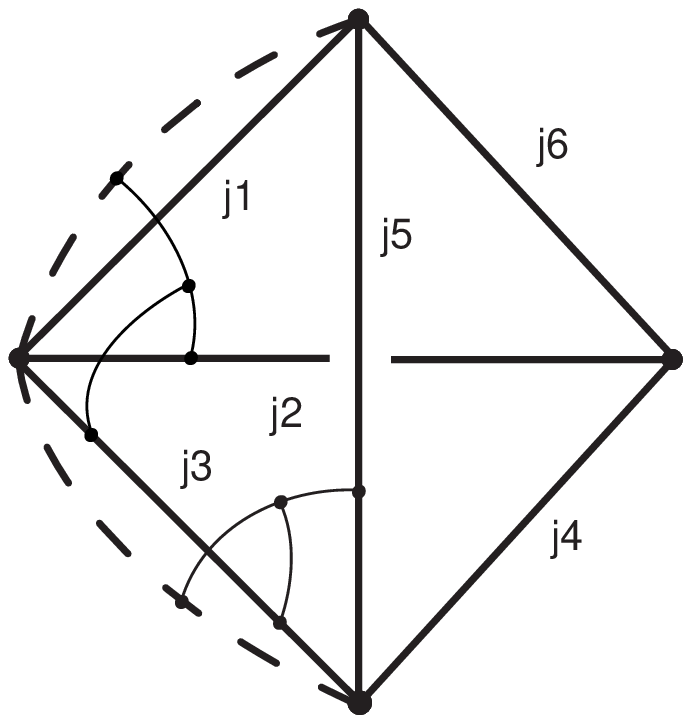} &
\psfrag{j1}{$j_1$}
\psfrag{j2}{$j_2$}
\psfrag{j3}{$j_3$}
\psfrag{j4}{$j_4$}
\psfrag{j5}{$j_5$}
\psfrag{j6}{$j_6$}
\includegraphics[scale=0.35]{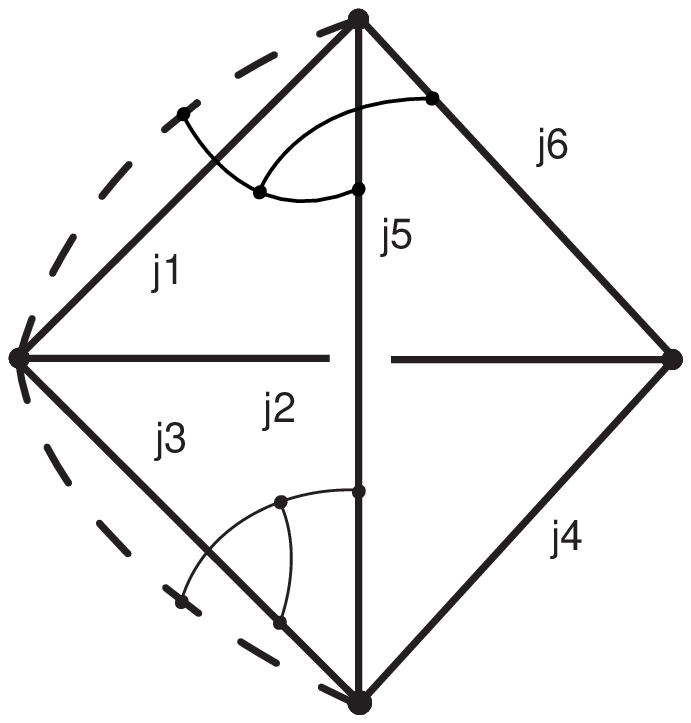} &
\psfrag{j1}{$j_1$}
\psfrag{j2}{$j_2$}
\psfrag{j3}{$j_3$}
\psfrag{j4}{$j_4$}
\psfrag{j5}{$j_5$}
\psfrag{j6}{$j_6$}
\includegraphics[scale=0.35]{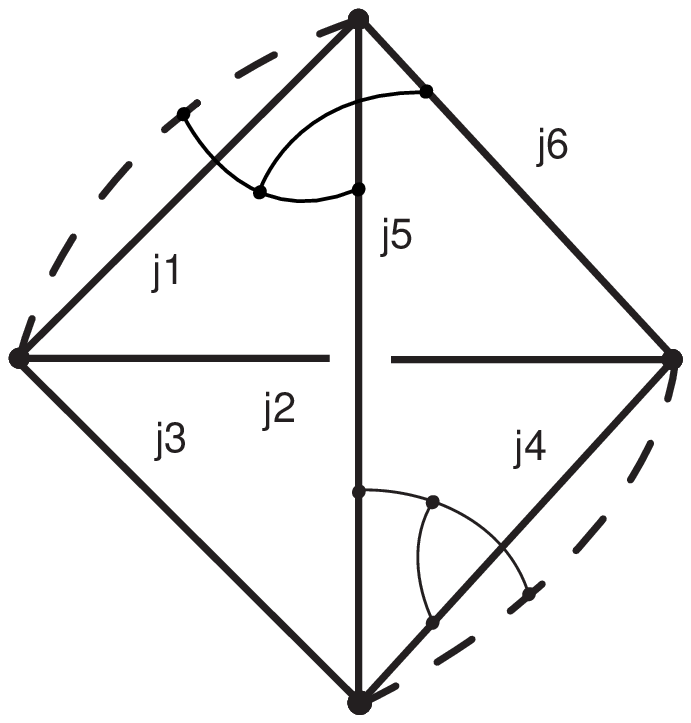} \\
  &
\psfrag{j1}{$j_1$}
\psfrag{j2}{$j_2$}
\psfrag{j3}{$j_3$}
\psfrag{j4}{$j_4$}
\psfrag{j5}{$j_5$}
\psfrag{j6}{$j_6$}
\includegraphics[scale=0.35]{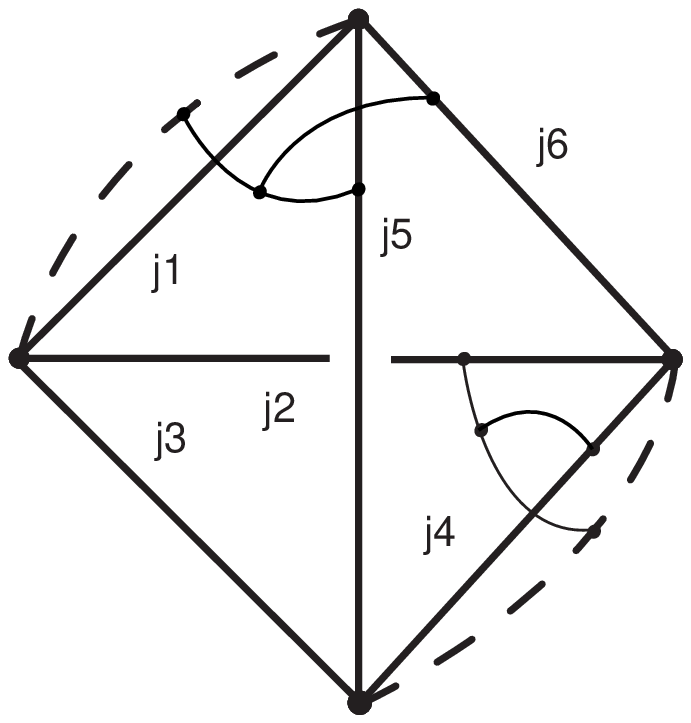}         &

\end{tabular}
\caption{Spin network diagrams for the different possible vertex amplitudes for the fermion spin foam model.  The dashed lines are in the spin half representation and the curved lines denote the grasping operators in the spin one representation.}
\label{fermion fig}
\end{figure}

Thus the partition function $Z_{\mbox{\tiny F}}(\Delta) $ for fermions coupled to 3d gravity can be considered as the expectation value of a particular observable for the usual GR partition function
\beqa
\label{Z fermions}
Z_{\mbox{\tiny F}}(\Delta) &=&
\sum_{\mathcal{L}_1 , \mathcal{L}_2}
\prod_{e^*} \int_{\SU(2)} dg_{e^*} \; \prod_{e'} \int_{{\mathfrak su}(2)}dX_{e'}
\;
\left( \prod_{\ell_i \in \mathcal{L}_1 } F_{\ell_i}     \right)
\left( \prod_{\ell_j \in \mathcal{L}_2 } F_{\ell_j}     \right)
\;
\exp \left( {{i} \sum_e {\rm Tr} \, X_e G_e } \right)
\eeqa
This can be computed with the generating functional techniques as for the volume operators.  One can see from  \eqref{Z fermions} that for each loop $\ell$, one obtains a Wilson loop in the spin half representation that is then grasped to the edges of each triangle it passes through.  Up to symmetry, the different vertex amplitudes for the resulting spin foam model are shown in figure \ref{fermion fig}.
Note that the orientation of the loops is important as the graspings act on the triangle corresponding to the point that the fermion enters the tetrahedron rather than the point that it leaves (neglecting the conjugate term in $F_\ell$).  In fact this is a simplification of the full fermion model in which the effect of the second term in \eqref{F loop} has been neglected but all of the important details are still preserved.
Again, in order to properly define the generating functional, one should use the wedge variables but the description given above should be clear enough for our purposes.

\subsubsection{ The quenched model}  In the calculation of a fermionic observable, an approximation can be made that neglects one of the fermionic loops, say $\mathcal{L}_2$.  This would correspond to considering a single symplectic majorana fermion in \cite{Fairbairn:2006dn}. There is then only one type of vertex amplitude
\beq
\mathcal{A}(\mathcal{L}_1 ) =
\begin{array}{c}
\psfrag{j1}{$j_1$}
\psfrag{j2}{$j_2$}
\psfrag{j3}{$j_3$}
\psfrag{j4}{$j_4$}
\psfrag{j5}{$j_5$}
\psfrag{j6}{$j_6$}
\psfrag{J}{$\frac{1}{2}$}
\includegraphics[scale=0.45]{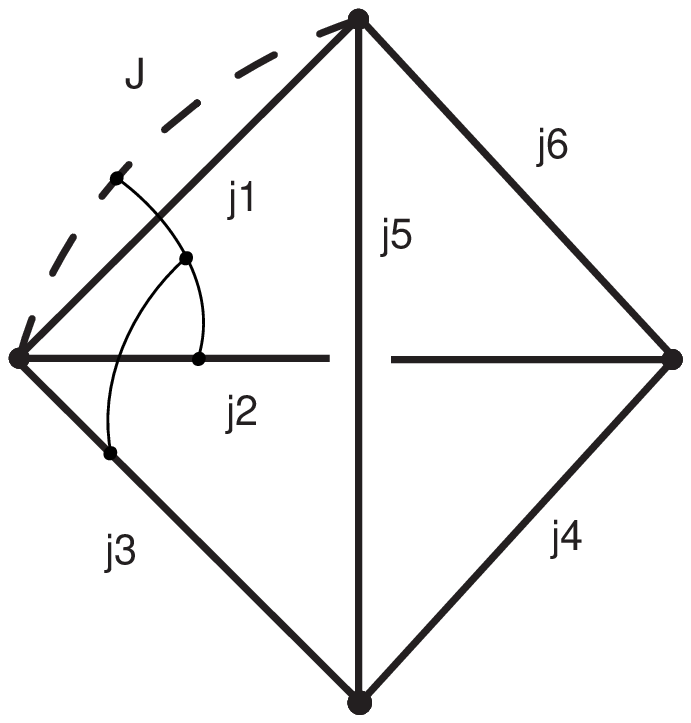}
\end{array}
\eeq
We can understand this as describing a fermion (the spin half line) entering a tetrahedron through one triangle, which is acted on by a grasping operator, then leaving through a different triangle.
For clarity we first describe the GFT for this model before moving onto the more complicated configurations possible when there are two fermionic loop configurations involved.
We will require three different fields to construct the vertex amplitude:
\begin{itemize}
\item The pure gravity field $\phi$ defined in Section \ref{wilson section} and obeying the same symmetries.
\beq
\label{GR field}
\phi(g_1,g_2,g_3) : SU(2) \times SU(2) \times SU(2) \rightarrow \mathbb{C}
\eeq
\item A Wilson loop field that has been projected to the spin half representation.
\beq
\psi^{\frac{1}{2}}(g_1,g_2,g_3 ; g) = (P^{\frac{1}{2}} \psi)(g_1,g_2,g_3 ; g)
: SU(2) \times SU(2) \times SU(2) \times SU(2) \rightarrow \mathbb{C}
\eeq
This corresponds to the fermion leaving the tetrahedron.
\item A final field that contains the grasping operator between the fermion line and the vertex at its source.  The field has three free spin one indices
\beq
\xi_{abc}(g_1,g_2,g_3 ; g)
: SU(2) \times SU(2) \times SU(2) \times SU(2)
\rightarrow \mathbb{V}_1  \otimes \mathbb{V}_1 \otimes \mathbb{V}_1,
\eeq
the fourth argument of the field is projected to the spin half representation
\beq
\xi^{\frac{1}{2}}_{abc}(g_1,g_2,g_3 ; g) = (P^{\frac{1}{2}} \xi_{abc})(g_1,g_2,g_3 ; g)
\eeq
and the following projection will provide the necessary graspings
\beqa
 (  \hat{P}_{ \beta_1  \beta_2  \beta_3 \alpha} \xi^{\frac{1}{2}}_{abc})(g_1,g_2,g_3 ; g)
 &=& \int_{SU(2)} \prod_{i=1}^3  d\beta_i d\alpha
  D^{1}_{ad}(\beta_1)  D^{1}_{be}(\beta_2) D^{1}_{cf}(\beta_3)
  \nn \\ && \times
  \xi^{\frac{1}{2}}_{def}(g_1 \alpha ,g_2 \beta_1 \alpha^{-1} \beta_1  ,g_3 \beta_2 \alpha^{-1} \beta_2 ; g \beta_3 \alpha^{-1} \beta_3)
\eeqa
We also allow the field to be unchanged under permutations $\sigma$ of the first three arguments
\beq
\xi_{abc}(g_1,g_2,g_3 ; g) =  \xi_{abc}(g_{\sigma(1)},g_{\sigma(2)},g_{\sigma(3)} ; g)
\eeq
This field describes the fermion entering the tetrahedron and grasping the appropriate triangle.
\end{itemize}
We can see that the action must now be constructed so that the fields appear in the correct configuration to describe a fermion entering and leaving the tetrahedron.
The mode expansions for $\phi, \psi^{\frac{1}{2}}$ are given above and the expansion for the quenched fermionic field $\xi_{abc}^{\frac{1}{2}}$ into modes $\left( \xi^{j_1 j_2 j_3 \frac{1}{2}}_{m_1 k_1 m_2 k_2 m_3 k_3 m_4 k_4}\right)_{def}$ gives
\beqa
 (  \hat{P}_{ \beta_1  \beta_2  \beta_3 \alpha} \xi^{\frac{1}{2}}_{abc})(g_1,g_2,g_3 ; g)
 &=&
 \sum_{ \substack{  j_i , m_i, n_i, q_3, a,b,c,s       \\    1 \leq i \leq 4       }       }
  \xi^{j_1 j_2 j_3 \frac{1}{2}}_{m_1 m_2 m_3 m_4}
  \sqrt{d_{j_1} d_{j_2}  d_{j_3} d_{\frac{1}{2}}}
    C^{j_2 j_2 1}_{n_2 q_2 a} C^{j_3 j_3 1}_{n_3 q_3 b} C^{\frac{1}{2} \frac{1}{2} 1}_{n_4 q_4 c}
     C^{j_1 j_2 j_3 \frac{1}{2} s }_{n_1 q_2 q_3 q_4}
\nn \\ && \times
D^{j_1}_{m_1 n_1}(g_1) D^{j_2}_{m_2 n_2}(g_2) D^{j_3}_{m_3 n_3}(g_3) D^{\frac{1}{2}}_{m_4 n_4}(g).
\eeqa
with the following redefinition of the modes
\beqa
\xi^{j_1 j_2 j_3 \frac{1}{2}}_{m_1 m_2 m_3 m_4}  =
\sum_{\substack{  k_i, p_2,p_3, p_4, d,e,f,t       \\    1 \leq i \leq 4       } }
\left( \xi^{j_1 j_2 j_3 \frac{1}{2}}_{m_1 k_1 m_2 k_2 m_3 k_3 m_4 k_4}\right)_{def}
 C^{j_2 j_2 1}_{p_2 k_2 d} C^{j_3 j_3 1}_{p_3 k_3 e} C^{\frac{1}{2} \frac{1}{2} 1}_{p_4 k_4 f}
     C^{j_1 j_2 j_3 \frac{1}{2} t }_{k_1 p_2 p_3 p_4}
.
\eeqa
The action for the theory is given by
\beqa
S_{\mbox{\tiny Q}}[\phi , \psi ,\xi, \lambda_{\mbox{\tiny Q}} ]
&=&
\frac{1}{2} \int \prod^3_{i=1} dg_i  dg \
P_{\alpha_1} \psi^{\frac{1}{2}}(g_1,g_2,g_3 ; g) \
P_{\alpha_2} \xi^{\frac{1}{2}}_{abc}(g_1,g_2,g_3 ; g) \
\epsilon_{abc}
\nn \\ && +
\frac{1}{2} \int \prod^3_{i=1} dg_i
P_{\alpha_1} \phi(g_1,g_2,g_3) \
P_{\alpha_2} \phi(g_1,g_2,g_3)
\nn \\ &&  +
\frac{\lambda_{\mbox{\tiny Q}}}{4} \int \prod^6_{i=1} dg_i dg \
(  \hat{P}_{ \beta_1  \beta_2  \beta_3 \alpha_1}
\xi^{\frac{1}{2}}_{abc})(g_1,g_2,g_3 ; g)  \
P_{\alpha_2} \psi^{\frac{1}{2}}(g_1,g_5,g_6 ; g)\
P_{\alpha_3} \phi(g_2,g_4,g_6) \
P_{\alpha_4} \phi(g_3,g_4,g_5)  \ \epsilon_{abc} \nn
\eeqa

\paragraph*{\bf Feynman rules:}
The partition function for the theory is
\beq
Z = \int  \mathcal{D} \psi^\frac{1}{2}   \D \phi
 \prod_{a,b,c=1}^{3} \D \xi^\frac{1}{2}_{abc}
e^{- S_{\mbox{\tiny Q}}[\phi, \psi^{\frac{1}{2}},\xi^\frac{1}{2}_{abc}, \lambda_{\mbox{\tiny Q}}]}
=\sum_{\Gamma}  \frac{  \lambda_{\mbox{\tiny Q}}^{v_{\mbox{\tiny Q}}[\Gamma ]}  }{\mathrm{sym}(\Gamma )} Z[\Gamma].
\eeq
As before, $v_{\mbox{\tiny Q}}[\Gamma ]$ denotes the number of vertices in the diagram.  In this case, the Feynman rules are more specific about the direction of propagation of the fermions throughout the 2-complex.  To represent this, we orient the fermion (dashed) lines and demand that the orientation is consistent within each diagram.  The propagators are given by
\begin{center}
\begin{tabular}{c}
\psfrag{g1}{$g_1$}
\psfrag{g2}{$g_2$}
\psfrag{g3}{$g_3$}
\includegraphics[scale=0.35]{grpropagator}
\end{tabular} \ \ \
\begin{tabular}{c}
\psfrag{g1}{$g_1$}
\psfrag{g2}{$g_2$}
\psfrag{g3}{$g_3$}
\psfrag{g}{$g$}
\includegraphics[scale=0.35]{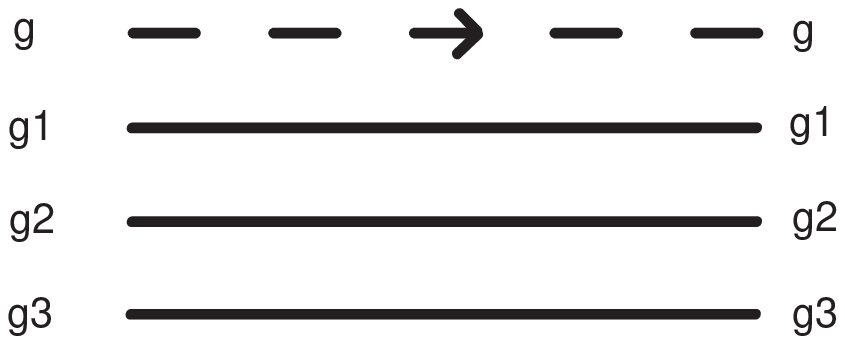}
\end{tabular}
\end{center}
The single vertex and corresponding amplitude of the quenched model is
\begin{center}
\psfrag{g1}{$g_1$}
\psfrag{g2}{$g_2$}
\psfrag{g3}{$g_3$}
\psfrag{g4}{$g_4$}
\psfrag{g5}{$g_5$}
\psfrag{g6}{$g_6$}
\psfrag{g}{$g$}
\begin{tabular}{c}
 \includegraphics[scale=0.4]{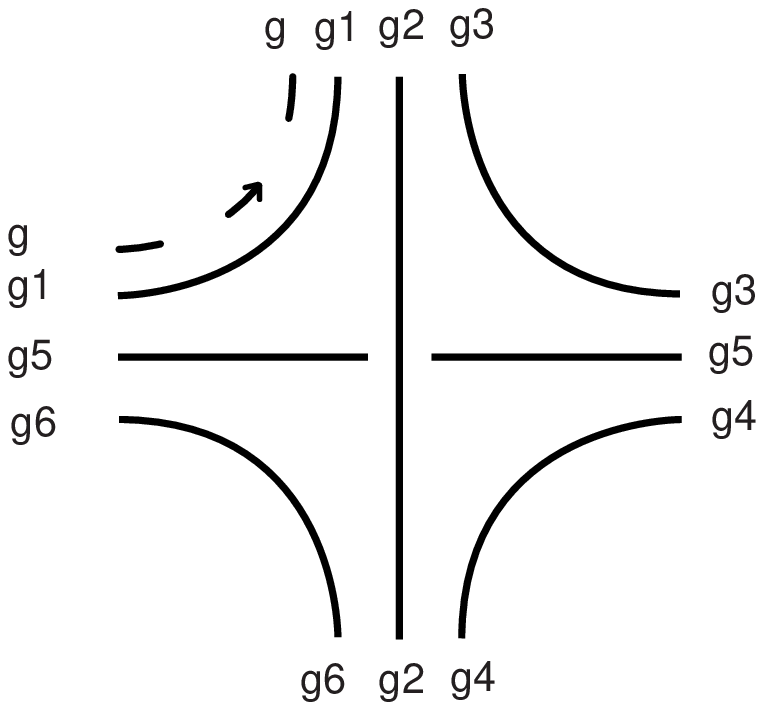}
 \end{tabular}
 { \Huge $\Rightarrow$}
 \begin{tabular}{c}
  \psfrag{j1}{$j_1$}
\psfrag{j2}{$j_2$}
\psfrag{j3}{$j_3$}
\psfrag{j4}{$j_4$}
\psfrag{j5}{$j_5$}
\psfrag{j6}{$j_6$}
\psfrag{J}{$\frac{1}{2}$}
\includegraphics[scale=0.45]{quenchedamp}
\end{tabular}
\end{center}

\subsubsection{The unquenched model}
We now construct the full GFT for massless fermions coupled to 3d gravity.  All of the important features of the model are contained in the quenched case, all that remains are the additional complications caused by the inclusion of two independent loop configurations instead of one.  As one can see from the fermionic vertex amplitudes illustrated above, we must provide a different interaction vertex for each possible configuration and orientation.  We will omit some of the detail as it is essentially the same as the quenched case and the complications involved are unenlightening.  We begin by defining the fields
\begin{itemize}
\item The pure gravity field $\phi$ defined in Section \ref{wilson section} and obeying the same symmetries.
\beq
\label{GR field 3}
\phi(g_1,g_2,g_3) : SU(2)^3 \rightarrow \mathbb{C}
\eeq
\item A Wilson loop field that has been projected to the spin half representation and corresponding to a fermion  leaving the tetrahedron through a triangle
\beq
\psi^{\frac{1}{2}}(g_1,g_2,g_3 ; g) = (P^{\frac{1}{2}} \psi)(g_1,g_2,g_3 ; g)
: SU(2)^4 \rightarrow \mathbb{C}
\eeq

\item A field containing two Wilson loops  corresponding to the ends of two fermion lines at a triangle
\beq
\psi_2(g_1,g_2,g_3 ; g , f ) =
 SU(2)^5 \rightarrow \mathbb{C}
\eeq
Both Wilson lines are projected to the spin half representation
\beq
\psi_2^{\frac{1}{2}, \frac{1}{2} }(g_1,g_2,g_3 ; g , f ) =
\int_{SU(2)} dh_1 \ d_{\frac{1}{2}} \chi^{\frac{1}{2}}(g h_1^{-1})
\int_{SU(2)} dh_2 \ d_{\frac{1}{2}} \chi^{\frac{1}{2}}(f h_2^{-1})
\psi_2(g_1,g_2,g_3 ; h_1,h_2),
\eeq
and the field has the usual invariance property by projection
\beq
P_\alpha \psi(g_1,g_2,g_3;g ,f) = \int_{SU(2)}  d\alpha \ \psi( g_1 \alpha, g_2 \alpha, g_3 \alpha ;  g \alpha, f \alpha)
\eeq

\item The fermion field $\xi_{abc}(g_1,g_2,g_3 ; g)$ described above which relates to a fermion entering a tetrahedron.

\item A tensor field that corresponds to two fermions passing through a triangle, one entering the tetrahedron and one leaving
\beq
\chi_{abc}(g_1,g_2,g_3 ; g , f )
: SU(2)^5
\rightarrow \mathbb{V}_1^{\otimes 3},
\eeq
The fourth and final arguments are projected to the spin half representation in the same way as $\psi_2^{\frac{1}{2}, \frac{1}{2} }$.  There will be a single grasping operation from the fermion entering the tetrahedron that will be provided by
\beqa
 (  \hat{P}_{ \beta_i  \alpha} \chi^{\frac{1}{2},\frac{1}{2}}_{abc})(g_1,g_2,g_3 ; g, f)
 &=& \int_{SU(2)} \prod_{i=1}^3  d\beta_i d\alpha
  D^{1}_{ad}(\beta_1)  D^{1}_{be}(\beta_2) D^{1}_{cf}(\beta_3)
  \nn \\ && \times
  \chi^{\frac{1}{2}, \frac{1}{2}}_{def}
  (g_1 \alpha ,g_2 \beta_1 \alpha^{-1} \beta_1  ,g_3 \beta_2 \alpha^{-1} \beta_2 ; g \beta_3 \alpha^{-1} \beta_3  ,f  \alpha)
\eeqa

\item A tensor field that describes two fermions entering the tetrahedron through the same triangle.
\beq
\tilde{\chi}_{abc,def}(g_1,g_2,g_3 ; g , f )
: SU(2)^5
\rightarrow   \mathbb{V}_1^{ \otimes 6}
\eeq
Again, the fourth and final arguments are projected to the spin half representation and we create the two grasping operators with
\beqa
 (  \tilde{P}_{ \beta_i  \gamma_i \alpha} \tilde{\chi}^{\frac{1}{2},\frac{1}{2}}_{abc,def})(g_1,g_2,g_3 ; g, f)
 &=& \int_{SU(2)} \prod_{i=1}^3  d\beta_i d\gamma_i d\alpha
  D^{1}_{ap}(\beta_1)  D^{1}_{bq}(\beta_2) D^{1}_{cr}(\beta_3)
  D^{1}_{ds}(\gamma_1)  D^{1}_{et}(\gamma_2) D^{1}_{fu}(\gamma_3)
  \nn \\ && \times
  \tilde{\chi}^{\frac{1}{2}, \frac{1}{2}}_{pqr,stu}
  (g_1 \gamma_1 \alpha^{-1} \gamma_1,
  g_2 \beta_1 \alpha^{-1} \beta_1  ,
  g_3 \gamma_2 \beta_2^{-1} \alpha \gamma_2^{-1} \beta_2 ;
  g \beta_3 \alpha^{-1} \beta_3  ,
  f  \gamma_3 \alpha^{-1} \gamma_3)
  \nn
\eeqa
This projection gives the field the following invariance property
\beqa
   D^{1}_{ap}(h)  D^{1}_{bq}(h) D^{1}_{cr}(h)
  D^{1}_{ds}(h)  D^{1}_{et}(h) D^{1}_{fu}(h)
   (  \tilde{P}_{ \beta_i  \gamma_i \alpha} \tilde{\chi}^{\frac{1}{2},\frac{1}{2}}_{pqr,stu})(g_1 h ,g_2 h ,g_3 h ; g h, f h)
   \nn \\
 = (  \tilde{P}_{ \beta_i  \gamma_i \alpha} \tilde{\chi}^{\frac{1}{2},\frac{1}{2}}_{abc,def})(g_1  ,g_2  ,g_3  ; g , f )
\eeqa
\end{itemize}
All of the new fields defined above are also invariant under permutations of the first three indices.
We can now write the action
\beqa
S_{\mbox{\tiny F}}
&=&
S_{\mbox{\tiny F}}[\phi , \psi^{\frac{1}{2}},\psi_2^{\half, \half} ,\xi^{\half}_{abc},\chi_{abc}^{\frac{1}{2},\frac{1}{2}},\tilde{\chi}^{\half, \half}_{abc,def}, \lambda_{\mbox{\tiny F}} ]
\nn \\ &=&
\frac{1}{2} \int \prod^3_{i=1} dg_i  dg \
P_{\alpha_1} \psi^{\frac{1}{2}}(g_1,g_2,g_3 ; g) \
P_{\alpha_2} \xi^{\frac{1}{2}}_{abc}(g_1,g_2,g_3 ; g) \
\epsilon_{abc}
\nn \\ && +
\frac{1}{2} \int \prod^3_{i=1} dg_i \
P_{\alpha_1} \phi(g_1,g_2,g_3) \
P_{\alpha_2} \phi(g_1,g_2,g_3)
\nn \\ &&  +
\frac{1}{2} \int \prod^3_{i=1} dg_i dg dh \
\hat{P}_{\beta_i  \alpha_1} \chi_{abc}^{\frac{1}{2},\frac{1}{2}}(g_1,g_2,g_3;g,h) \
\hat{P}_{\beta'_i \alpha_2} \chi_{def}^{\frac{1}{2},\frac{1}{2}}(g_1,g_2,g_3;g,h)
\epsilon_{abc} \epsilon_{def}
\nn \\ &&  +
\frac{1}{2} \int \prod^3_{i=1} dg_i dg dh \
\tilde{P}_{\beta_i \gamma_j \alpha_1} \tilde{\chi}_{abc,def}^{\frac{1}{2},\frac{1}{2}} (g_1,g_2,g_3;g,h) \
P_{\alpha_2} \psi_2^{\frac{1}{2},\frac{1}{2}} (g_1,g_2,g_3;g,h)
\epsilon_{abc} \epsilon_{def}
\nn \\ &&  +
\frac{\lambda_{\mbox{\tiny F}}}{4} \int \prod^6_{i=1} dg_i dg dh \
\tilde{P}_{\beta_i \gamma_j \alpha_1} \tilde{\chi}_{abc,def}^{\frac{1}{2},\frac{1}{2}} (g_1,g_2,g_3;g,h) \
P_{\alpha_2} \psi_2^{\frac{1}{2},\frac{1}{2}} (g_1,g_5,g_6;g,h)
P_{\alpha_3} \phi(g_2,g_4,g_6) \
P_{\alpha_4} \phi(g_3,g_4,g_5)
\epsilon_{abc} \epsilon_{def}
\nn \\ &&  +
\frac{\lambda_{\mbox{\tiny F}}}{4} \int \prod^6_{i=1} dg_i
\hat{P}_{\beta_i \alpha_1} \chi_{abc}^{\frac{1}{2},\frac{1}{2}} (g_1,g_2,g_3;g,h) \
\hat{P}_{\beta'_j \alpha_2} \chi_{def}^{\frac{1}{2},\frac{1}{2}} (g_1,g_5,g_6;g,h)
P_{\alpha_3} \phi(g_2,g_4,g_6) \
P_{\alpha_4} \phi(g_3,g_4,g_5)
\epsilon_{abc} \epsilon_{def}
\nn \\ &&  +
\frac{\lambda_{\mbox{\tiny F}}}{4} \int \prod^6_{i=1} dg_i dg dh \
\tilde{P}_{\beta_i \gamma_j \alpha_1} \tilde{\chi}_{abc,def}^{\frac{1}{2},\frac{1}{2}} (g_1,g_2,g_3;g,h) \
P_{\alpha_2} \psi^{\frac{1}{2}} (g_1,g_5,g_6;g)
P_{\alpha_3} \psi^{\frac{1}{2}} (g_3,g_4,g_5;h) \
P_{\alpha_4} \phi(g_2,g_4,g_6)
\epsilon_{abc} \epsilon_{def}
\nn \\ &&  +
\frac{\lambda_{\mbox{\tiny F}}}{4} \int \prod^6_{i=1} dg_i dg dh \
\hat{P}_{\beta_i \alpha_1} \chi_{abc}^{\frac{1}{2},\frac{1}{2}} (g_1,g_2,g_3;g,h)
\hat{P}_{\alpha_2} \xi^{\frac{1}{2}}_{abc}(g_1,g_5,g_6 ; g)
P_{\alpha_3} \psi^{\frac{1}{2}} (g_3,g_4,g_5;h)
 P_{\alpha_4} \phi(g_2,g_4,g_6)
 \epsilon_{abc} \epsilon_{def}
\nn \\ &&  +
\frac{\lambda_{\mbox{\tiny F}}}{4} \int \prod^6_{i=1} dg_i dg dh \
P_{\alpha_1} \psi_2^{\frac{1}{2},\frac{1}{2}} (g_1,g_2,g_3;g,h)
\hat{P}_{\alpha_2} \xi^{\frac{1}{2}}_{abc}(g_1,g_5,g_6 ; g)
\hat{P}_{\alpha_3} \xi^{\frac{1}{2}}_{abc}(g_2,g_4,g_6 ; g)
P_{\alpha_4} \phi(g_3,g_4,g_5)
\epsilon_{abc} \epsilon_{def}
\nn \\ &&  +
\frac{\lambda_{\mbox{\tiny F}}}{4} \int \prod^6_{i=1} dg_i dg dh \
\hat{P}_{\alpha_1} \xi^{\frac{1}{2}}_{abc}(g_1,g_2,g_3 ; g)
\hat{P}_{\alpha_2} \xi^{\frac{1}{2}}_{abc}(g_1,g_5,g_6 ; h)
P_{\alpha_3} \psi^{\frac{1}{2}} (g_2,g_4,g_6;g)
P_{\alpha_4} \psi^{\frac{1}{2}} (g_3,g_4,g_5;h)
\epsilon_{abc} \epsilon_{def}
\eeqa

\paragraph*{\bf Feynman rules:}
The partition function for the theory is
\beq
Z = \int
\D \phi
\D \psi^\half
\D \psi^{\half,\half}_2
\prod_{a,...,m=1}^{3}
\D \xi^\frac{1}{2}_{abc}
\D \chi^{\half \half}_{def}
\D \tilde{\chi}^{\half \half}_{ghi,klm}
e^{- S_{\mbox{\tiny F}}}
=\sum_{\Gamma}  \frac{  \lambda_{\mbox{\tiny F}}^{v_{\mbox{\tiny F}}[\Gamma ]}  }{\mathrm{sym}(\Gamma )} Z[\Gamma].
\eeq
The four types of propagator are (in the order given in the action)
\begin{center}
\psfrag{g1}{$g_1$}
\psfrag{g2}{$g_2$}
\psfrag{g3}{$g_3$}
\psfrag{g4}{$g_4$}
\psfrag{g5}{$g_5$}
\psfrag{g6}{$g_6$}
\psfrag{g}{$g$}
\psfrag{h}{$h$}
\includegraphics[scale=0.35]{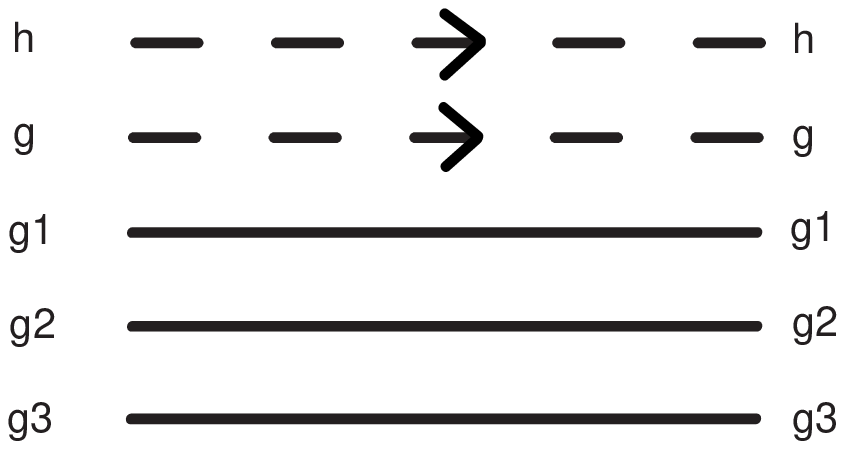}  \ \ \
\psfrag{g1}{$g_1$}
\psfrag{g2}{$g_2$}
\psfrag{g3}{$g_3$}
\psfrag{g4}{$g_4$}
\psfrag{g5}{$g_5$}
\psfrag{g6}{$g_6$}
\psfrag{g}{$g$}
\includegraphics[scale=0.35]{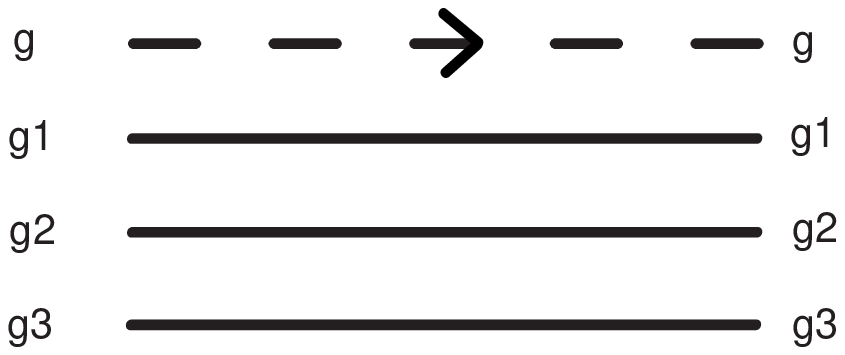} \ \ \
\psfrag{g1}{$g_1$}
\psfrag{g2}{$g_2$}
\psfrag{g3}{$g_3$}
\psfrag{g4}{$g_4$}
\psfrag{g5}{$g_5$}
\psfrag{g6}{$g_6$}
\psfrag{g}{$g$}
\psfrag{h}{$h$}
\includegraphics[scale=0.35]{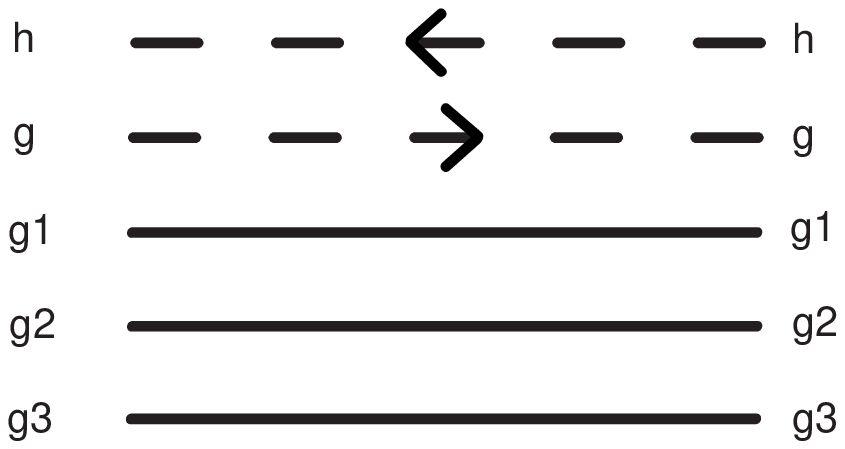}  \ \ \
\psfrag{g1}{$g_1$}
\psfrag{g2}{$g_2$}
\psfrag{g3}{$g_3$}
\psfrag{g4}{$g_4$}
\psfrag{g5}{$g_5$}
\psfrag{g6}{$g_6$}
\includegraphics[scale=0.35]{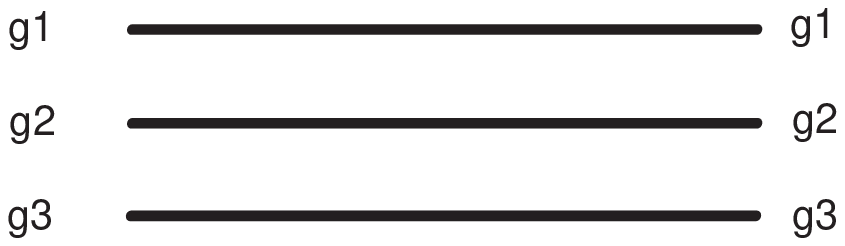}
\end{center}
The vertices, listed in the order given in the action, and their corresponding vertex amplitudes are

\begin{center}

\begin{tabular}{c}
\psfrag{g1}{$g_1$}
\psfrag{g2}{$g_2$}
\psfrag{g3}{$g_3$}
\psfrag{g4}{$g_4$}
\psfrag{g5}{$g_5$}
\psfrag{g6}{$g_6$}
\psfrag{g}{$g$}
\psfrag{h}{$h$}
 \includegraphics[scale=0.45]{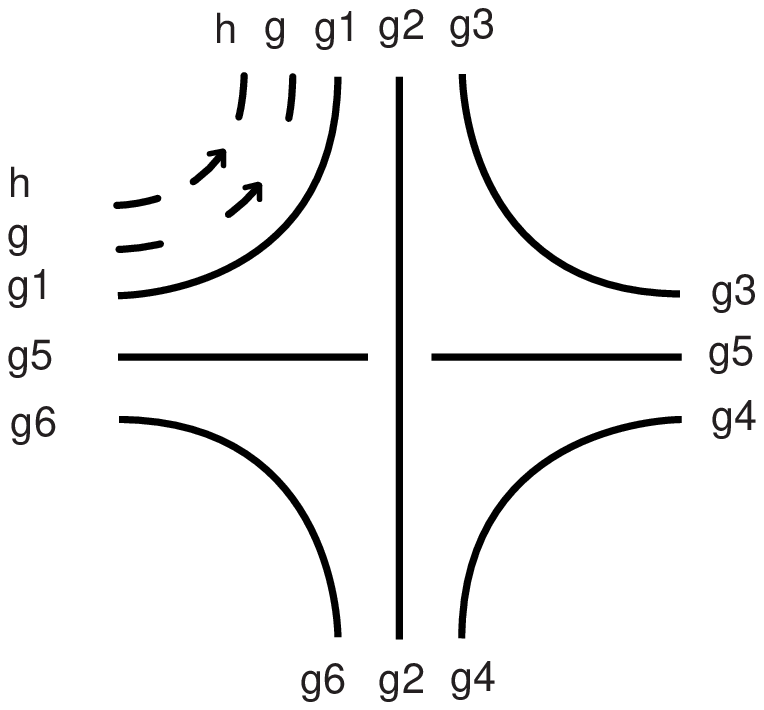}
 \end{tabular}
 { \Huge $\Rightarrow$}
 \begin{tabular}{c}
   \psfrag{j1}{$j_1$}
\psfrag{j2}{$j_2$}
\psfrag{j3}{$j_3$}
\psfrag{j4}{$j_4$}
\psfrag{j5}{$j_5$}
\psfrag{j6}{$j_6$}
\psfrag{J}{$\frac{1}{2}$}
\includegraphics[scale=0.45]{chipsiphiphiamp}
\end{tabular} ,
\begin{tabular}{c}
\psfrag{g1}{$g_1$}
\psfrag{g2}{$g_2$}
\psfrag{g3}{$g_3$}
\psfrag{g4}{$g_4$}
\psfrag{g5}{$g_5$}
\psfrag{g6}{$g_6$}
\psfrag{g}{$g$}
\psfrag{h}{$h$}
 \includegraphics[scale=0.45]{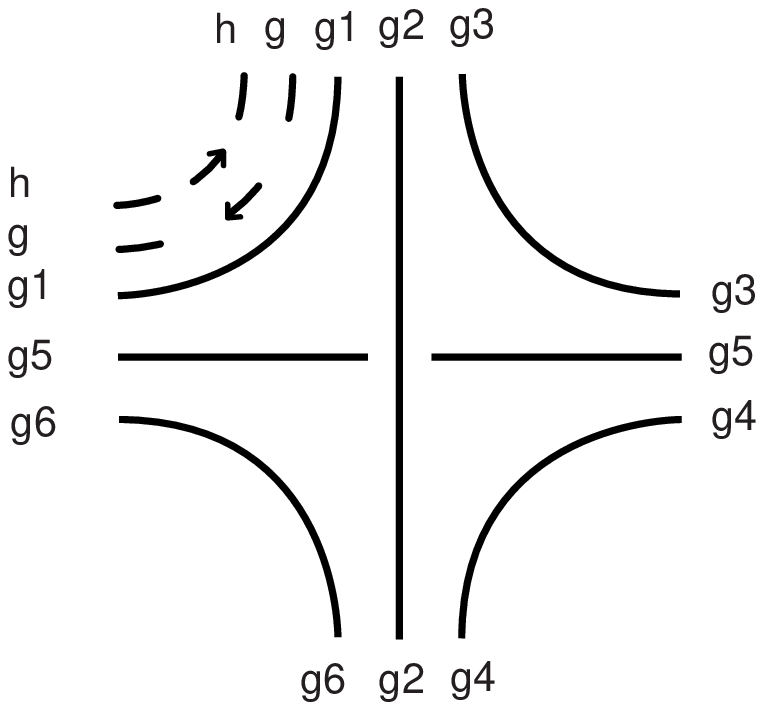}
 \end{tabular}
 { \Huge $\Rightarrow$}
 \begin{tabular}{c}
    \psfrag{j1}{$j_1$}
\psfrag{j2}{$j_2$}
\psfrag{j3}{$j_3$}
\psfrag{j4}{$j_4$}
\psfrag{j5}{$j_5$}
\psfrag{j6}{$j_6$}
\psfrag{J}{$\frac{1}{2}$}
\includegraphics[scale=0.45]{chichiphiphiamp}
\end{tabular} \\

\begin{tabular}{c}
\psfrag{g1}{$g_1$}
\psfrag{g2}{$g_2$}
\psfrag{g3}{$g_3$}
\psfrag{g4}{$g_4$}
\psfrag{g5}{$g_5$}
\psfrag{g6}{$g_6$}
\psfrag{g}{$g$}
\psfrag{h}{$h$}
 \includegraphics[scale=0.45]{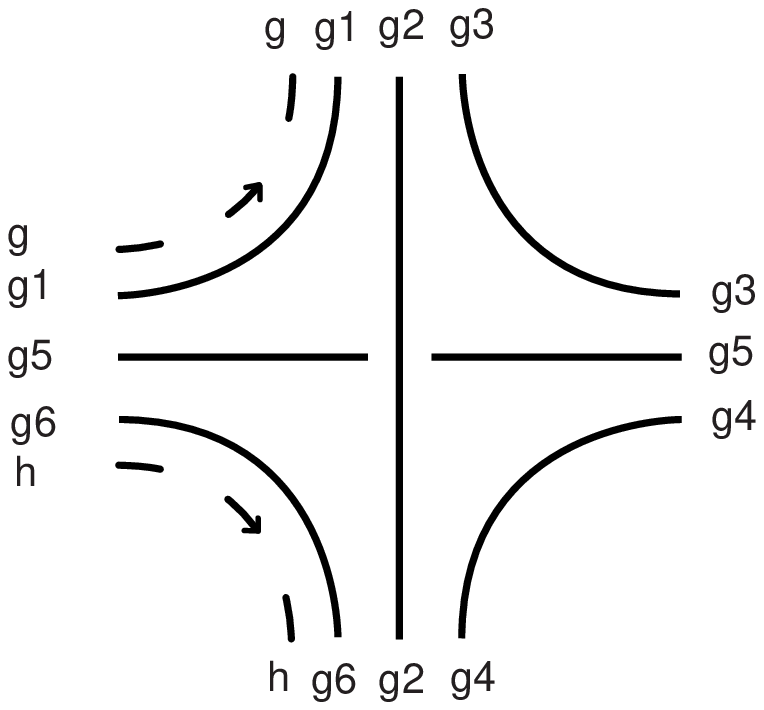}
 \end{tabular}
 { \Huge $\Rightarrow$}
 \begin{tabular}{c}
    \psfrag{j1}{$j_1$}
\psfrag{j2}{$j_2$}
\psfrag{j3}{$j_3$}
\psfrag{j4}{$j_4$}
\psfrag{j5}{$j_5$}
\psfrag{j6}{$j_6$}
\psfrag{J}{$\frac{1}{2}$}
\includegraphics[scale=0.45]{chipsipsiphiamp}
\end{tabular} ,
\begin{tabular}{c}
\psfrag{g1}{$g_1$}
\psfrag{g2}{$g_2$}
\psfrag{g3}{$g_3$}
\psfrag{g4}{$g_4$}
\psfrag{g5}{$g_5$}
\psfrag{g6}{$g_6$}
\psfrag{g}{$g$}
\psfrag{h}{$h$}
 \includegraphics[scale=0.45]{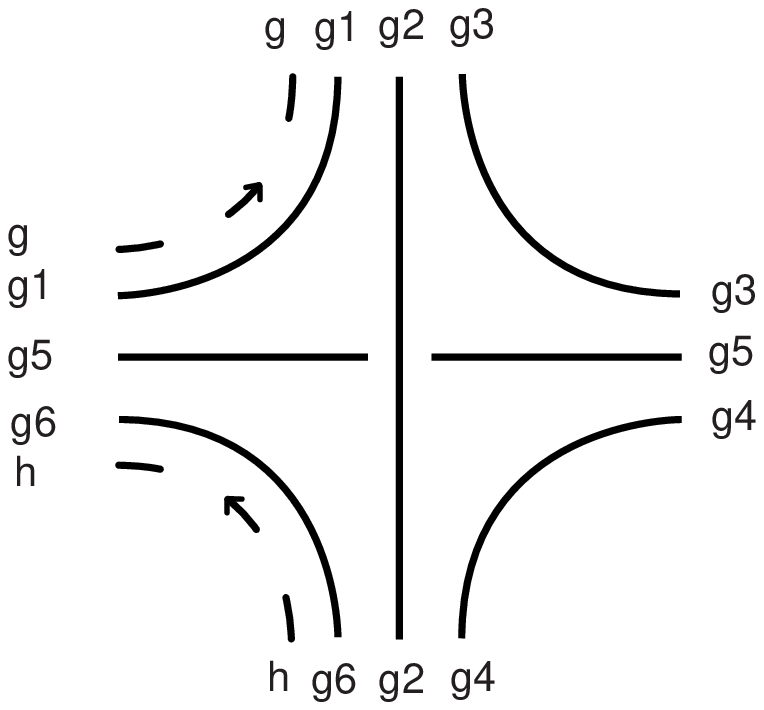}
 \end{tabular}
 { \Huge $\Rightarrow$}
 \begin{tabular}{c}
    \psfrag{j1}{$j_1$}
\psfrag{j2}{$j_2$}
\psfrag{j3}{$j_3$}
\psfrag{j4}{$j_4$}
\psfrag{j5}{$j_5$}
\psfrag{j6}{$j_6$}
\psfrag{J}{$\frac{1}{2}$}
\includegraphics[scale=0.45]{chixipsiphiamp}
\end{tabular} \\

\begin{tabular}{c}
\psfrag{g1}{$g_1$}
\psfrag{g2}{$g_2$}
\psfrag{g3}{$g_3$}
\psfrag{g4}{$g_4$}
\psfrag{g5}{$g_5$}
\psfrag{g6}{$g_6$}
\psfrag{g}{$g$}
\psfrag{h}{$h$}
 \includegraphics[scale=0.45]{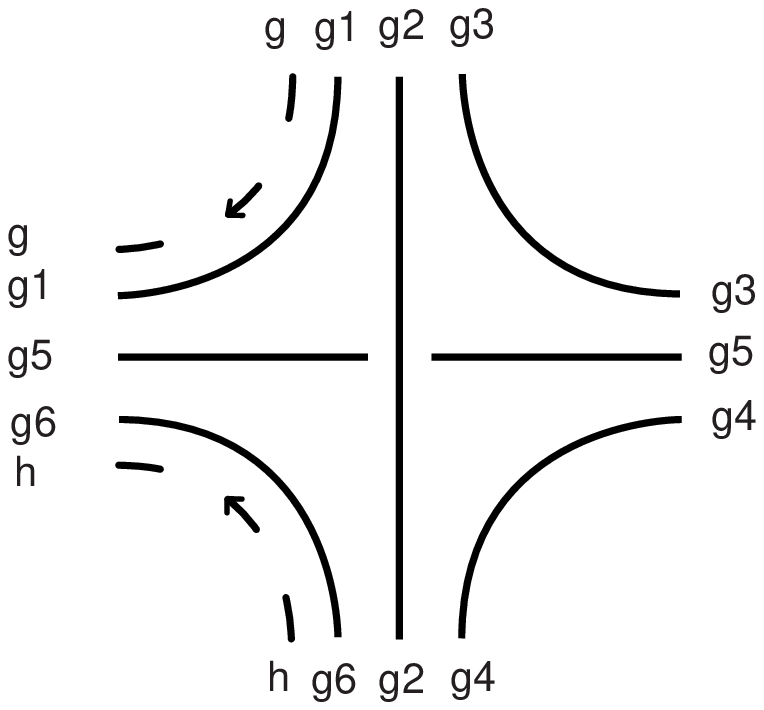}
 \end{tabular}
 { \Huge $\Rightarrow$}
 \begin{tabular}{c}
    \psfrag{j1}{$j_1$}
\psfrag{j2}{$j_2$}
\psfrag{j3}{$j_3$}
\psfrag{j4}{$j_4$}
\psfrag{j5}{$j_5$}
\psfrag{j6}{$j_6$}
\psfrag{J}{$\frac{1}{2}$}
\includegraphics[scale=0.45]{psixixiphiamp}
\end{tabular},
\begin{tabular}{c}
\psfrag{g1}{$g_1$}
\psfrag{g2}{$g_2$}
\psfrag{g3}{$g_3$}
\psfrag{g4}{$g_4$}
\psfrag{g5}{$g_5$}
\psfrag{g6}{$g_6$}
\psfrag{g}{$g$}
\psfrag{h}{$h$}
 \includegraphics[scale=0.45]{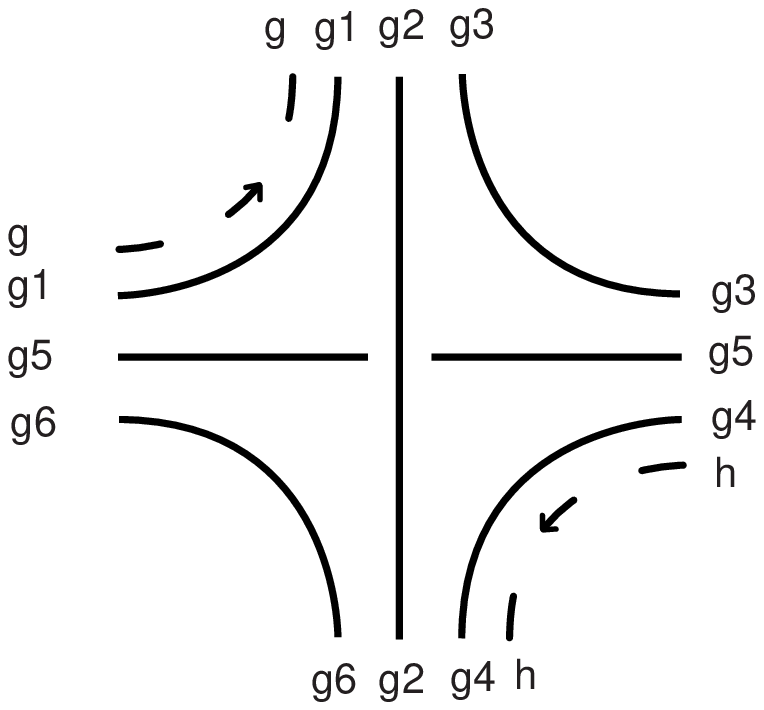}
 \end{tabular}
 { \Huge $\Rightarrow$}
 \begin{tabular}{c}
    \psfrag{j1}{$j_1$}
\psfrag{j2}{$j_2$}
\psfrag{j3}{$j_3$}
\psfrag{j4}{$j_4$}
\psfrag{j5}{$j_5$}
\psfrag{j6}{$j_6$}
\psfrag{J}{$\frac{1}{2}$}
\includegraphics[scale=0.45]{xixipsipsiamp}
\end{tabular} \\

\begin{tabular}{c}
\psfrag{g1}{$g_1$}
\psfrag{g2}{$g_2$}
\psfrag{g3}{$g_3$}
\psfrag{g4}{$g_4$}
\psfrag{g5}{$g_5$}
\psfrag{g6}{$g_6$}
\psfrag{g}{$g$}
\psfrag{h}{$h$}
 \includegraphics[scale=0.45]{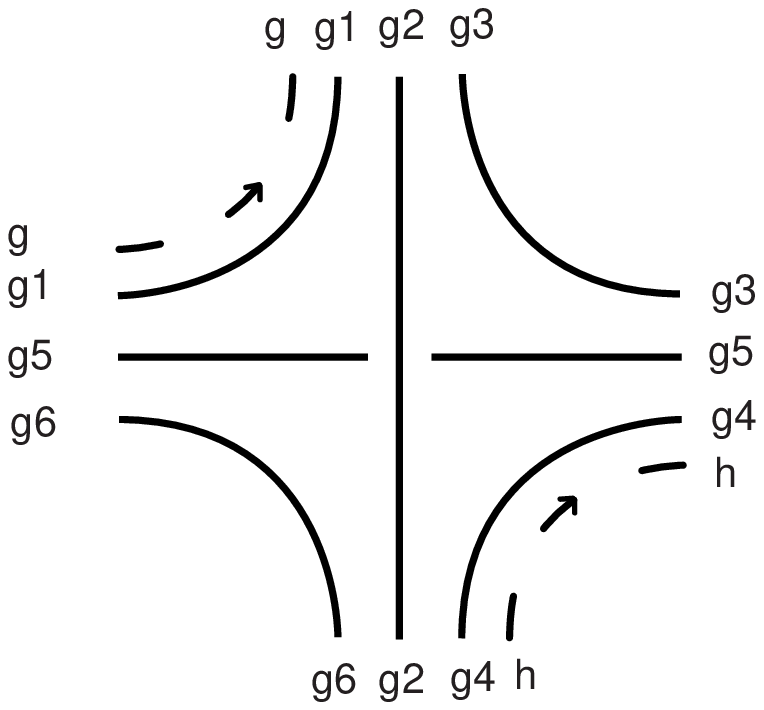}
 \end{tabular}
 { \Huge $\Rightarrow$}
 \begin{tabular}{c}
    \psfrag{j1}{$j_1$}
\psfrag{j2}{$j_2$}
\psfrag{j3}{$j_3$}
\psfrag{j4}{$j_4$}
\psfrag{j5}{$j_5$}
\psfrag{j6}{$j_6$}
\psfrag{J}{$\frac{1}{2}$}
\includegraphics[scale=0.45]{xixipsipsiamp2}
\end{tabular}

\end{center}

The Feynman diagrams for this GFT reproduce the spin foam model for fermions coupled to 3d gravity defined in \cite{Fairbairn:2006dn} with the simplification mentioned previously and the relevant numerical factors absorbed into the coupling coefficient.
Note that the fermion spin foam model is only defined on a proper triangulation, not any 2-complex.  Some of the Feynman diagrams of the  GFT will therefore produce spin foams that do not relate to the original definition of the model.  For example, those in which  two adjacent tetrahedra are glued on more than one face and the fermions propagate though both of these faces. However, in this particular degenerate example, one can show that the amplitude is zero.

We can also now see that at each value of $v_{\mbox{\tiny F}}[\Gamma ]$ we obtain all possible loop configurations that completely saturate the spin foam.  Thus the GFT naturally includes the matter diagrams as well as the sum over 2-complexes.  This will of course include all matter diagrams on all topologies available for a given value of $v_{\mbox{\tiny F}}[\Gamma ]$.

\section{Conclusions}

We have altered the standard GFT for 3d gravity to include observables in the form of Wilson loops, volume operators and fermions.  While the physical significance of summing over Wilson loops and volume operators is unclear, they are both essential for constructing the fermionic GFT.

The fermionic GFT gives a way to deal with the triangulation dependence in the spin foam model for fermions, one of the issues mentioned in \cite{Fairbairn:2006dn} and, as one would expect, implements the sum over fermionic loops at the same time.
We now briefly point out the differences between the fermionic GFT and the GFT for point particles.  Firstly, the corresponding spin foam models have very different vertex amplitudes, even for the spinning particle. The particles live on the edges of the spin foam since they have an associated deficit angle but the fermions propagate along the dual edges as they make use of the $\SU(2)$ variables to parallel transport the spinors.
Secondly, the particle model allows 3 and 4-valent interaction vertices whereas these are not allowed in the fermionic spin foam due to the Grassmann integration.
In the fermionic model, there are two separate configurations of fermionic loops and the loops must necessarily be closed and saturate all of the vertices (again due to the Grassmann variables) whereas there is only a single arbitrary graph in the particle model.  Modifications could of course be made to the interaction terms in the particle model to remove some of these differences and considering fermionic observables would change the allowed amplitudes of the fermion model.

It would be desirable to give an alternative GFT for the fermion model which takes several Grassmann variables as its arguments.  In this way it may be possible to recreate the model with fewer interaction vertices, however, it would also need to include a dependence on the frame fields in a similar way to \cite{Oriti:2007vf}.  Note that this suggestion is not the same as that in \cite{Gurau:2009tw} where the fields themselves are Grassmann valued.

\bibliographystyle{hieeetr}
\bibliography{Bibliography2009}

\end{document}